%% file: hd45166_paper2_vfin.tex
\documentclass{aa}
\usepackage{natbib}
\usepackage{aalongtable}
\bibpunct{(}{)}{;}{a}{}{,}
\usepackage{graphicx}
\usepackage{txfonts}

\newcommand{\kms}{\ensuremath{{\rm km\,s^{-1}}}}                   
\newcommand{\msun}{\ensuremath{\mathrm{M}_{\odot}}}                  
\newcommand{\msunyr}{\ensuremath{\mathrm{M}_{\odot}{\rm yr}^{-1}}}   
\newcommand{\lsun}{\ensuremath{\mathrm{L}_{\odot}}}                  
\newcommand{\rsun}{\ensuremath{\mathrm{R}_{\odot}}}                  

\newcommand{\lstar}{\ensuremath{\mathit{L}_{\star}}}                 
\newcommand{\mdot}{\ensuremath{\dot{M}}}                             
\newcommand{\mstar}{\ensuremath{\mathit{M}_{\star}}}                 
\newcommand{\rstar}{\ensuremath{\mathit{R}_{\star}}}                 
\newcommand{\teff}{\ensuremath{\mathit{T}_{\rm eff}}}                
\newcommand{\reff}{\ensuremath{\mathit{R}_{\rm phot}}}                
\newcommand{\vinf}{\ensuremath{v_{\infty}}}                          
\newcommand{\vesc}{\ensuremath{v_{\rm esc}}}                         
\newcommand{\tstar}{\ensuremath{\mathit{T}_{\star}}}                 
\newcommand{\K}{\ensuremath{\mathrm{K}}}                             

\newcommand{\vrot}{\ensuremath{v_{\rm rot}}}                         
\newcommand{\vcrit}{\ensuremath{v_{\rm crit}}}                         
\newcommand{\prot}{\ensuremath{P_{\rm rot}}}                         
\newcommand{\pcrit}{\ensuremath{P_{\rm crit}}}                         

\begin{document}

\title{The qWR star HD~45166\thanks{Based on observations made with the 1.52 m
ESO telescope at La Silla, Chile.}}
\subtitle{ II. Fundamental stellar parameters and evidence of a latitude-dependent wind}

\author{J. H. Groh \inst{1,2}\fnmsep \thanks{ \email{jgroh@mpifr-bonn.mpg.de}} \and A. S. Oliveira \inst{3,4} \and J. E. Steiner
\inst{2}}

\institute{Max-Planck-Institute f\"ur Radioastronomie, Auf dem H\"ugel 69, D-53121 Bonn, Germany \and 
Instituto de Astronomia, Geof\'{i}sica e Ci\^encias Atmosf\'ericas,
Universidade de S\~ao Paulo, Rua do Mat\~ao 1226, Cidade Universit\'aria, 05508-900,
S\~ao Paulo, SP, Brazil \and IP\&D, Universidade do Vale do Para\'{i}ba, Av. Shishima Hifumi, 2911, CEP 12244-000, S\~ao Jos\'e
dos Campos, SP, Brazil \and  SOAR Telescope, Casilla 603, La Serena, Chile}

\authorrunning{J. H. Groh et al.}
\titlerunning{The latitude-dependent wind of HD~45166}

\date{Received  / Accepted }

\abstract{The enigmatic object HD~45166 is a qWR
star in a binary system with an orbital period of 1.596 day, and presents a rich emission-line spectrum in addition to
absorption lines from the companion star (B7~V). As the system inclination is very small ($i=0.77 ^{\circ} \pm 0.09 ^{\circ}$),
HD~45166 is an ideal laboratory for wind-structure studies. }{The goal of the present paper is to determine
the fundamental stellar and wind parameters of the qWR star.}{A radiative transfer model for the wind and photosphere of the qWR
star was calculated using the non-LTE code CMFGEN. The wind asymmetry was also analyzed using a recently-developed version 
of CMFGEN to compute the emerging spectrum in two-dimensional geometry. The temporal-variance spectrum (TVS) was
calculated for studying the line-profile variations.}{Abundances,
stellar and wind parameters of the qWR star were obtained. The qWR star has an effective temperature of $\teff = 50000
\pm 2000$~\K, a luminosity of $\mathrm{log} (L/\lsun) = 3.75 \pm 0.08$, and a corresponding photospheric radius of $\reff
= 1.00$~\rsun. The star is helium-rich (N(H)/N(He) = 2.0), while the CNO abundances are anomalous when compared either
to solar values, to planetary nebulae, or to WR stars. The mass-loss rate is $\mdot
= 2.2 \times 10^{-7}$~\msunyr, and  the wind terminal velocity is $\vinf=425$~\kms. The comparison between the observed line profiles and
models computed under different latitude-dependent wind
densities strongly suggests the presence of an oblate wind density enhancement, with a density contrast of at least 8:1 from
equator to pole. If a high velocity polar wind is present ($\sim 1200$~\kms), the minimum density contrast is reduced
to 4:1.}{The wind parameters determined are unusual when compared to O-type stars or to typical WR stars. While for WR
stars $\vinf/\vesc > 1.5$, in the case of HD~45166 it is much smaller ($\vinf/\vesc = 0.32$). In addition, the efficiency of momentum
transfer is $\eta=0.74$, which is at least 4 times smaller than in a typical WR. We find evidence for the presence of
a wind compression zone, since the equatorial wind density is significantly larger when
compared to the polar wind. The TVS supports the presence of such a latitude-dependent wind and a variable absorption/scattering gas near the equator.}
\keywords{Stars: winds, outflows - Stars: mass-loss - Stars: fundamental parameters - binaries: spectroscopic - Stars: individual: HD~45166 - Stars: Wolf-Rayet}
\maketitle

\section{\label{intro}Introduction}

HD~45166 has been observed since 1922 \citep{anger33}, without much advancement in the understanding of its nature.
\citet{vb78} analyzed the optical H and He lines assuming that HD~45166 is a Population I WR object, and concluded
that the WR component has a radius of 1~\rsun, and has a small-sized envelope expanding with a  velocity of 150~\kms. 
The resulting number density of He II is about $10^{11}$~cm$^{-3}$ and mimics the environment of a WR
envelope. He found a mass-loss rate of $4.5 \times 10^{-8}$~\msunyr, and wind densities of N(He) = $3 \times
10^{11}$~cm$^{-3}$  and N(H)=$10^{12}$~cm$^{-3}$. Analyzing the IUE data, \citet{ws83} derived $\mathrm{log}
(L/\lsun) = 3.84$, a radius of $R=0.77$~\rsun, and an effective temperature of \teff=60000~K. \citet{wsh89} obtained a
wind terminal velocity of 1200~\kms, derived from the UV resonance lines. Recently, \citet{willis06} analyzed the far-ultraviolet spectrum
of HD~45166. By fitting the continuum energy distribution from the far ultraviolet to the near-infrared, they derived that the qWR star has $R=1.20$~\rsun, $\mathrm{log} (L/\lsun) = 3.28$, and $\teff=35000$~K. A mass-loss rate of $2 \times 10^{-7}$~\msunyr~was inferred by those authors using the optical lines of \ion{He}{ii}, \ion{C}{iii}, and \ion{N}{iii}.

\citet{so05} (hereafter Paper I) showed that HD~45166 is a double-lined binary system composed of a qWR and a
B7~V star in a system with an orbital period of 1.596 day. HD~45166 presents a rich emission-line spectrum in addition to
the absorption spectrum due to the cooler component. The orbital parameters of the system were derived in Paper I, showing
that the orbit is slightly eccentric ($e=0.18 \pm 0.08$) and has a very small inclination angle ($i=0.77 ^{\circ} \pm 0.09 ^{\circ}$). The
masses are $M_1=4.2$~\msun~and $M_2=4.8$~\msun. In addition to the orbital period, two other
periods were found in the qWR star (5 and 15 hours, Paper I). As the system inclination is very small, HD~45166 is an interesting laboratory
for studying the wind structure. 

The goal of the present paper is to study in detail the stellar and wind parameters of the qWR star. For this
purpose we use the radiative transfer code CMFGEN \citep{hm98} to analyze the high-resolution optical spectrum
presented in Paper I. The temporal variance spectrum (hereafter TVS; \citealt{fgb96}) of the strongest emission lines is also analyzed in order to obtain insights on the wind structure.

This paper is organized as follows. In Sect.~\ref{obs}, we briefly summarize the data presented in Paper~I, and
describe how the spectrum of the qWR star was disentangled from the B7~V companion. In Sect.~\ref{model}, we present
the main characteristics of CMFGEN, while in Sect.~\ref{res} the results of the quantitative analysis using the
spherical modeling are shown. The latitude dependence of the wind is analyzed in Sect.~\ref{2d}, using a
recently-developed version of CMFGEN \citep{bh05}. The analysis of the TVS is presented in Sect.~\ref{tvs}, and the
results obtained in this work are discussed in Sect.~\ref{discussion}, especially the
presence of a wind-compression zone. Finally, Sect.~\ref{conc} summarizes the main conclusions of this paper.

\section{\label{obs}Observations and subtraction of the B7~V companion spectrum}

The data analyzed on this paper were taken in January 2004 with the Fiber-fed Extended Range Optical
Spectrograph (FEROS, \citealt{kaufer99}) at the 1.52~m telescope of the European Southern Observatory (ESO) in La
Silla, Chile. The spectra have a resolution power of R=48\,000, and were reduced using the standard data-reduction
pipeline \citep{stahl99}. A total of 40 spectra with individual exposure times of 15~min were averaged, resulting
in a total integration time of 10~h. 

Since our goal in this paper is to obtain the fundamental parameters of the qWR star, its spectrum has first to be
disentangled from the spectrum of the B7~V companion. This task was accomplished by:
\begin{enumerate}
\item flux calibrating the average observed spectrum of HD~45166 using the available photometry (de-reddened);
\item scaling the flux of a standard synthetic B7~V continuum spectrum, obtained from \citet{pickles98}, to a distance of 1.3~kpc
(Paper I);
\item multiplying the normalized spectrum of the B7~V star HD~62714 obtained from the UVES Paranal Observatory Project \citep{bagnulo03}
by the Pickles B7~V scaled continuum obtained above, to account for the spectral features of the B7~V companion;
\item subtracting this scaled B7~V spectrum from the flux-calibrated observed spectrum of HD~45166;
\item normalizing the resultant spectrum by the continuum, which can then be directly compared to the
continuum-normalized model spectrum.
\end{enumerate}

\begin{figure}
\resizebox{\hsize}{!}{\includegraphics{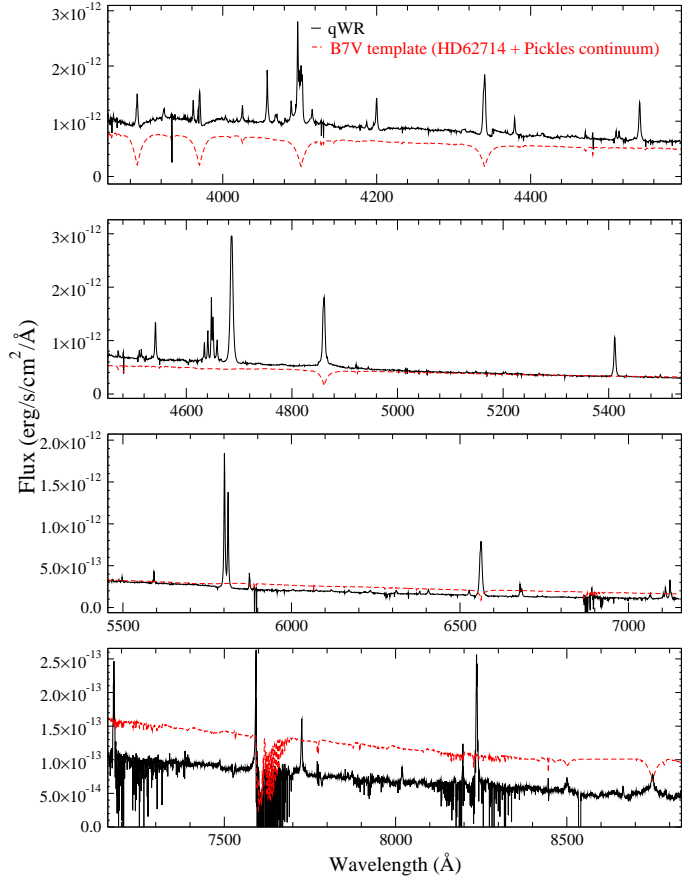}} 
\caption{\label{specobs} Resulting de-reddened spectrum of HD~45166 decomposed into the qWR star (full black line) and the B7~V
companion (dashed red line). As anticipated in Paper I, the qWR star is dominant in the blue part of the spectrum, while the B7~V
companion dominates for  $\lambda>5000 {\AA}$.}
\end{figure}

In general, this procedure results in a very good subtraction of the continuum and reasonable subtraction of the hydrogen absorption
lines due to the B7~V star, as can be seen in Fig.~\ref{specobs}. However, the rest of the rich absorption-line spectrum of the companion star is
not subtracted perfectly, and weak residual lines can still be seen. Nevertheless, their presence do not affect the diagnostic
lines used for obtaining the parameters of the qWR, as the companion lines are quite narrow, weak, and often are not
blended with the lines of the qWR star.

\section{\label{model}The Model}

We used the radiative transfer code CMFGEN \citep{hm98} to analyze the spectrum of the hot component of HD~45166 in
detail. The code assumes a spherically-symmetric outflow in steady-state, computing line and continuum formation in
the non-LTE regime. Each model is specified by the hydrostatic core radius \rstar, the luminosity \lstar, the
mass-loss rate \mdot, the wind terminal velocity \vinf, and the chemical abundances $Z_i$ of the included species.
Since the code does not solve the momentum equation of the wind, a velocity law must be adopted. The velocity
structure $v(r)$ is parameterized by a beta-type law, which is modified at depth to smoothly match a hydrostatic
structure at $\rstar$ (defined at a Rosseland optical depth of $\tau=100$ in our models). {\it The use of a hydrostatic structure at depth is crucial to analyze HD~45166, since the continuum and some of the higher ionization lines are formed in this region.}

\begin{table}
\caption{Final atomic model used in the analysis of HD~45166. A single-level ion of one higher ionization stage is included for every species (e.g. \ion{C}{vi} for carbon) and omitted below for brevity.}  
\label{modatom} 
\centerline{
\begin{tabular}{l r r r}
\hline\hline 
Ion & $N_\mathrm{S}$ & $N_\mathrm{F}$ & $N_\mathrm{T}$\\
\hline
\ion{H}{i}    &  30 & 20 & 435 \\
\ion{He}{i}   &  45 & 40 & 435 \\
\ion{He}{ii}  &  30 & 22 & 233 \\
\ion{C}{iii}  &  243 & 83 & 5513 \\
\ion{C}{iv}   &  64 & 60 & 1446 \\
\ion{C}{v}    &  107 & 43 & 2196 \\
\ion{N}{iii}  &  287 & 57 & 344 \\
\ion{N}{iv}   &  60 & 34 & 331 \\
\ion{N}{v}    &  67 & 45 & 1554 \\
\ion{O}{iii}  &  349 & 64 & 823 \\
\ion{O}{iv}   &  95 & 19 & 823 \\
\ion{O}{v}    &  34 & 34 & 118 \\
\ion{O}{vi}    &  13 & 13 & 59 \\
\ion{Si}{iv}	&  33 & 22 & 185 \\
\ion{Si}{v}	&  45 & 22 & 302 \\
\ion{Fe}{iv}    & 264 & 28 & 3889 \\
\ion{Fe}{v}	& 191 & 47 & 2342 \\
\ion{Fe}{vi}	& 433 & 44 & 8663 \\
\ion{Fe}{vii}	& 252 & 41 & 2734 \\
\ion{Fe}{viii}	& 324 & 53 & 7305 \\
\ion{Ni}{iv}	& 200 & 36 & 2337 \\
\ion{Ni}{v}	& 183 & 46 & 1524 \\
\hline
\end{tabular}}
\par \smallskip $N_\mathrm{S}$ = number of included superlevels; $N_\mathrm{F}$ = number of total energy\protect levels; $N_\mathrm{T}$ = number of bound-bound
transitions included for each ion.  
\end{table} 

The code includes the effects of clumping via a volume filling factor $f$ which depends on the distance following
an exponential law that supposes an unclumped wind when $r = \rstar$. The clumps start to form at the distance where
$v=v_c$, and the wind reaches its maximum clumping at large $r$:
\begin{equation}
f(r)=f+(1-f)\exp[-v(r)/v_c]\,\,.
\end{equation}

CMFGEN includes directly the influence of line blanketing in the ionization structure of the wind and,
consequently, in the calculated spectrum. With the concept of super-levels, the equations of statistical
equilibrium and radiative transfer can be solved simultaneously including thousands of lines in NLTE. The atomic
model used for HD~45166 consists of 35059 spectral lines from 2426 energy levels grouped in 617 super-levels of
H, He, C, N, O, Si, Fe, and Ni, as shown in Table \ref{modatom}.

\section{\label{res}Results: fundamental parameters of the qWR star}

\subsection{Effective temperature, luminosity and radius}

The effective temperature was constrained using the relative strength of lines from different ionization
stages of He, C, and N. The diagnostic lines used to obtain the He ionization structure were \ion{He}{i} $\lambda$
5876, \ion{He}{i} $\lambda$ 6678, \ion{He}{ii} $\lambda$ 4686, and \ion{He}{ii} $\lambda$ 5411, while for the C
ionization structure we used \ion{C}{iii} $\lambda$$\lambda$ 4647--4650--4651 and \ion{C}{iv}
$\lambda$$\lambda$ 5801--5812. The N ionization structure was obtained using the lines of \ion{N}{iii} $\lambda$
4097, \ion{N}{iii} $\lambda$ 4634, \ion{N}{iv} $\lambda$ 4058, \ion{N}{iv} $\lambda$$\lambda$ 7103--7109--7123, 
\ion{N}{v} $\lambda$ 4605, \ion{N}{v} $\lambda$ 4620, and \ion{N}{v} $\lambda$ 4945.

\begin{figure}
\resizebox{\hsize}{!}{\includegraphics{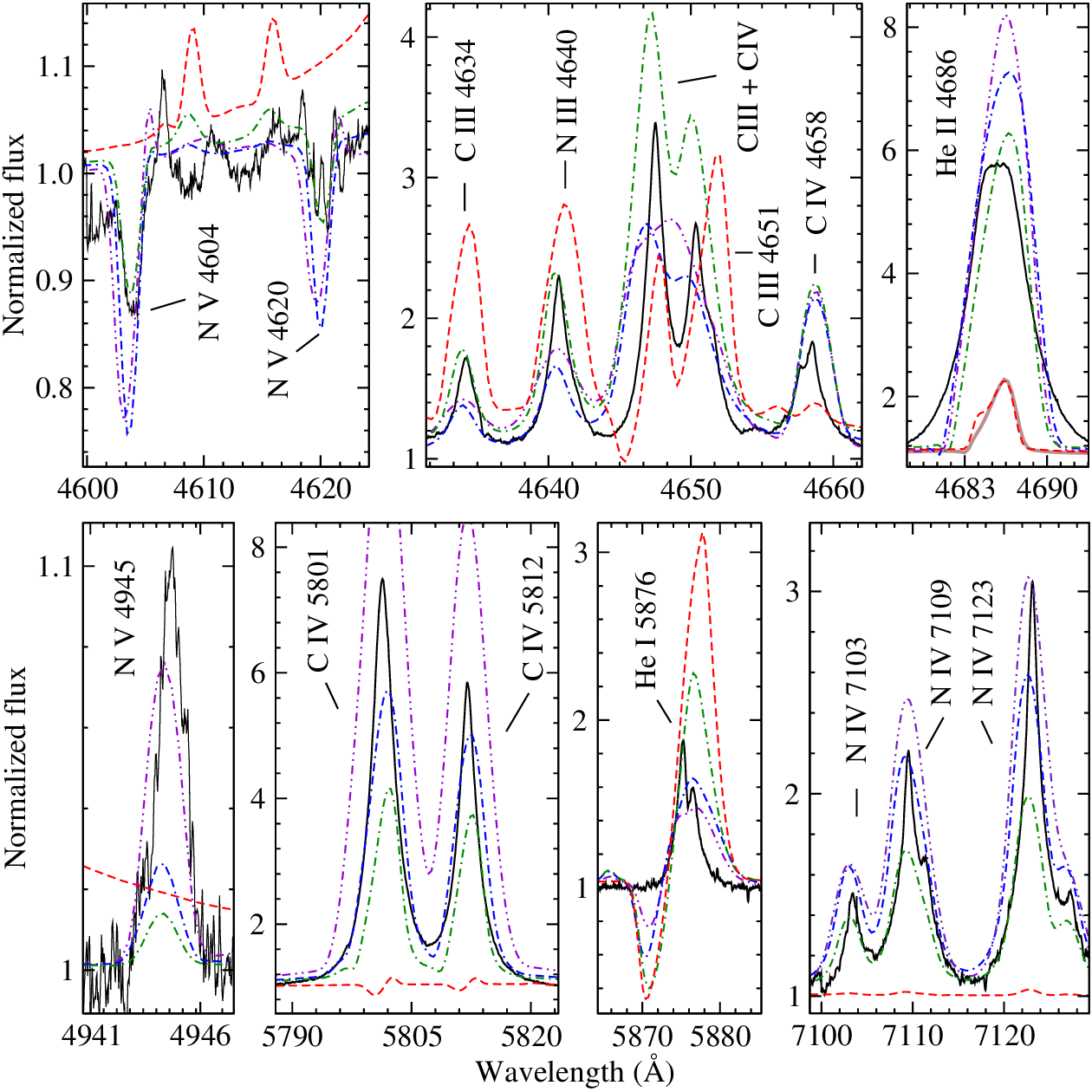}} 
\caption{\label{sepfitsteff} Normalized spectrum of the qWR star in HD~45166 (black line) compared with spherical CMFGEN models with $\teff=50\,000~\mathrm{K}$ (purple dash-double-dotted line), $\teff=46\,000~\mathrm{K}$ (blue dot-double-dashed line), $\teff=43\,000~\mathrm{K}$ (green dot-dashed line),  and $\teff=35\,000~\mathrm{K}$ (red dashed line). A model with the same parameters proposed by \citet{willis06} ($\teff=35\,000~\mathrm{K}$, $\vinf=1200~\kms$) produces fits similar as the original $\teff=35\,000~\mathrm{K}$ model, and does not fit our optical data as well. For clarity, such a model is compared only to \ion{He}{ii} $\lambda$ 4686 (brown thick line). }
\end{figure}

Figure \ref{sepfitsteff} presents the fits to the diagnostic lines used to derive $\teff$ in this work. In particular, the relative strength between \ion{He}{i} and \ion{He}{ii} lines requires $\teff > 47\,000~\mathrm{K}$, otherwise the \ion{He}{i} lines in the model become too strong compared to the observations, and the \ion{He}{ii} lines become too weak. The N ionization structure requires models with $\teff > 46\,000~\mathrm{K}$ in order to reproduce the ratio between the \ion{N}{iii} and \ion{N}{iv} lines mentioned above. Reasonable fits to the \ion{N}{v} lines are only achieved by models with $\teff > 50\,000~\mathrm{K}$. The strongest C lines seen in the spectrum, namely \ion{C}{iv} $\lambda$$\lambda$ 5801--5812, also require models with $\teff > 49\,000~\mathrm{K}$ to obtain reasonable fits to the observations. CMFGEN models with $\teff < 48\,000~\mathrm{K}$ produce too strong \ion{C}{iii} emission. 

As can be seen in Fig. \ref{sepfitsteff}, the fits to the observed line spectrum are very sensitive to the effective temperature, and the use of three diagnostics in this temperature regime allowed us to constrain the effective temperature as $\teff(\tau=2/3)=50\,000 \pm 2000$~K. The temperature of the hydrostatic core was constrained to $\tstar(\tau=100)=70\,000 \pm 2000$~\K and obtained through the use of a hydrostatic structure at depth. The hydrostatic structure is sensitive to the effective gravity and, hence, to the adopted stellar mass ($\mathrm{M}=4.2~\msun$, Paper I). 

It is worthwhile noting that even if it was not possible to reproduce lines corresponding to the observed ionization stages of He, C, and N using a unique best model with $\teff=50\,000~\mathrm{K}$, small changes in the range $\pm 2000~~\mathrm{K}$ were sufficient to adjust the discrepant lines (Fig. \ref{sepfitsteff}). 

Models with other parameter regimes do not fit the optical spectrum of HD~45166 from 2004. For instance, increasing the terminal velocity of a model with $\teff=35\,000~\mathrm{K}$ to $\vinf=1200~\kms$ (i.e., the same model parameters of \citealt{willis06}) does not enhance the ionization degree of the wind and, therefore, does not fit the high-ionization optical lines. Such high-terminal velocity model provides fits similar to the original $\teff=35\,000~\mathrm{K}$ model with $\vinf=425~\kms$ (Fig. \ref{sepfitsteff}).

The fundamental parameters of the qWR star obtained in this work, using tailored CMFGEN models to fit the continuum energy distribution {\it and } the optical spectrum, differ significantly from the previous published values. While we cannot exclude variability, the discrepancies between our results and previous works are at least partially explained by the inclusion of key physical ingredients in our analysis; namely, the inclusion of detailed non-LTE radiative transfer in the co-moving frame, full line blanketing, and use of a hydrostatic structure at depth, which allows simultaneous modeling of the photosphere and wind. However, since \citet{willis06} used the same radiative transfer code as ours, the aforementioned physical ingredients were presumably included in their modeling as well. The value of $\teff$ inferred by those authors (35\,000~K) is 15\,000~K lower than our determination, and this difference is puzzling. 

\begin{figure}
\resizebox{\hsize}{!}{\includegraphics{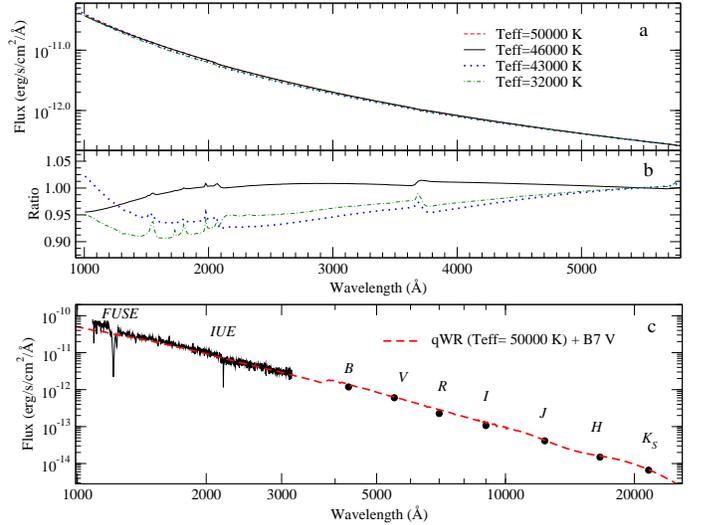}} 
\caption{\label{modflux} {\it a)} Comparison between the continuum flux predicted by CMFGEN models for the qWR star in the range $32\,000 \mathrm{K}< \teff < 50\,000 
\mathrm{K}$. All models were scaled to the observed de-reddened flux of the qWR star at 5500 {\AA} ($3.09 \times 106{-13}$ erg s$^{-1}$ cm$^{-2}$ {\AA} $^{-1}$, see Fig. \ref{specobs}). Since the continuum slope is very weakly sensitive on the value of $\teff$ (see text for discussion), the models are basically superimposed to each other.  {\it b)} Ratio between the continuum flux predicted by a CMFGEN model with a given $\teff$ and the continuum flux of the $\teff = 50\, 000$ K CMFGEN model.  {\it c)} Comparison between the total de-reddened observed flux of HD~45166 (black) with the flux predicted by the radiative-transfer model (red dashed line). The model shown here is composed by the sum of the continuum of the B7~V template from the companion (shown in Fig. \ref{specobs}) and the continuum of the best CMFGEN model for the qWR star.}
\end{figure}

In order to investigate this discrepancy in the values of $\teff$, we computed several CMFGEN models covering a broad range of effective temperatures in order to analyze the changes in the continuum, since this was the diagnostic of the effective temperature used by \citet{willis06}. While the spectral lines are strongly sensitive to changes in $\teff$ (Fig. \ref{sepfitsteff}), we found that the continuum slope predicted by CMFGEN models in the range $32\,000~\mathrm{K}< \teff < 50\,000~\mathrm{K}$ is almost insensitive to the adopted value of $\teff$ (Fig. \ref{modflux}a,b). Indeed, all such models provide good fits to the ultraviolet-to-near-infrared SED of HD~45166 (Fig. \ref{modflux}c). This is not surprising since such hot stars emit the bulk of their flux in the range 228--1000 {\AA}. Similar results have long been found for other hot stars, such as Wolf-Rayet stars \citep{hillier87,abbott87} and O-type stars \citep{martins06}. We also computed models using a large value of the wind terminal velocity, as proposed by \citet{willis06}, and found no noticeable changes in the continuum slope. This happens because the photosphere is located at rather low velocities ($\sim 30~\kms$) in comparison with Wolf-Rayet stars, and changes in the velocity field have little impact on the continuum formation region and in the continuum spectrum shown in Fig. \ref{modflux}. 

In the case of HD~45166, only small changes (of the order of 8\% or less) are seen in the ultraviolet when comparing models in the range $32\,000~\mathrm{K}< \teff < 50\,000~\mathrm{K}$ (Fig. \ref{modflux}b). Those changes are easily compensated by slightly adjusting other model parameters, such as the mass-loss rate, core radius, or reddening law. Even without changing any parameter, those small changes in the ultraviolet flux tend to be masked by observational errors, since the $1\sigma$ error in the absolute photometry of the IUE and FUSE flux-calibrated spectrum is at least 5\%. {\it Therefore, since models in the range $32\,000~\mathrm{K}< \teff < 50\,000~\mathrm{K}$ provide good fits to the observed SED (Fig. \ref{modflux}c),  we suggest that the technique of fitting the continuum is not suitable for constraining $\teff$ in the case of HD ~45166, and very likely explains the discrepancy between the low $\teff$ suggested by \citet{willis06} and the value determined in this work.}

The stellar luminosity of the qWR star was obtained by matching the best-model flux, scaled to a distance of $d=1.3$ kpc (Paper
I), with the observed flux of HD~45166, de-reddened using $E(B-V)$=0.155. The photometry was taken from \citet{wsh89},
from Paper I, and references therein. We obtained a luminosity of $\lstar=5650$~$\lsun$ ($\mathrm{log} (L/\lsun) =
3.75$), with an error due to the uncertainties in the modeling, distance, reddening law, and photometry amounting
to about 20\% (0.08 dex). Figure~\ref{modflux}c displays the de-reddened observed flux compared with the flux from the
best CMFGEN model. We determined a luminosity significantly higher than that given by \citet{willis06}, which is due to the significantly higher $\teff$ inferred by our detailed modeling of the spectral lines. 

Combining the derived values of $\tstar$ and $\teff$ with \lstar, and using the Stefan-Boltzmann law, it is possible to
determine the radius of the hydrostatic core of the qWR star as $\rstar(\tau=100)=0.51~$\rsun, and the radius of the
photosphere (defined as $\tau=2/3$) as $\reff=1.00~$\rsun.

\subsection{The mass-loss rate}

We derived $\mdot = 2.2 \times 10^{-7}$~$\msunyr$ in order to reproduce the strength of the spectral lines present in the optical
spectrum taken in 2004. We also obtained a volume filling factor of $f=0.5$, using the electron-scattering wings of 
\ion{He}{ii} $\lambda$ 4686 and \ion{C}{iv} $\lambda$$\lambda$ 5801--5812 as diagnostics. However, the electron scattering
wings are weak, and we cannot rule out the presence of an unclumped wind ($f=1$), with $\mdot = 3.1 \times 10^{-7}$~\msunyr.  
The value of the clumped $\mdot$ derived in this work agrees with the proposed value given by \citet{willis06}, while the value of the unclumped 
$\mdot$ derived here is 50\% higher than the value obtained by them.

Using the value of the mass-loss rate combined with the velocity law derived in Sect \ref{vinf} and the equation of mass
continuity, we obtain the wind density structure shown in Fig.~\ref{velden}.

\subsection{\label{vinf}Wind terminal velocity and momentum transfer}

\begin{figure}
\resizebox{\hsize}{!}{\includegraphics{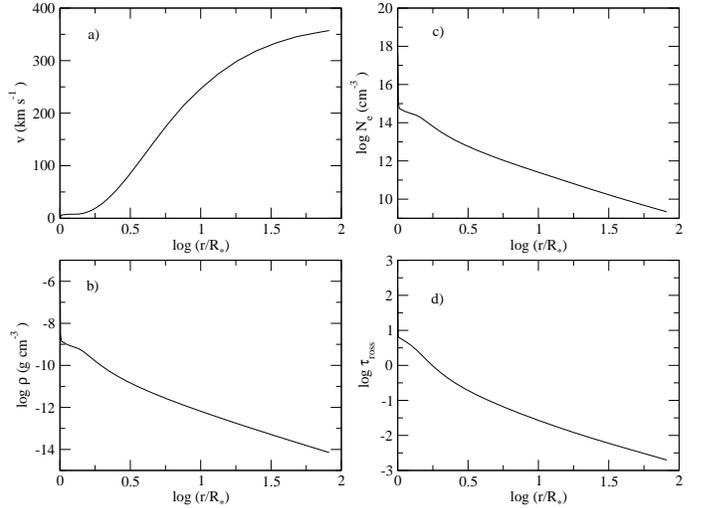}} 
\caption{\label{velden} Velocity law (panel $a$), density structure ($b$), electron density ($c$), and Rosseland
optical depth ($d$) of HD~45166 as a function of distance.}
\end{figure}

From our optical spectrum we did not detect velocities above $+550~\kms$ for the \ion{He}{ii} $\lambda$ 4686 line (see
Paper I). For the \ion{He}{i} $\lambda$ 5876 line, however, the maximum velocity observed is much smaller (280~\kms). For the
\ion{C}{iv} $\lambda$$\lambda$ 5801--5812 lines, a strong emission exists at low velocity, while weak emission can be seen
extending up to 600~\kms. From the model fit to \ion{He}{ii} $\lambda$ 4686 and \ion{He}{ii}+H$\alpha$ $\lambda$ 6560 , we
obtained a wind terminal velocity of $\vinf=425 \pm 50$~\kms. The value of the acceleration parameter $\beta$ was set to 4.0 in
order to reproduce the relative strength among the \ion{He}{ii} lines of the Pickering series. In particular, it was impossible to
reproduce the high-order \ion{He}{ii} lines of the Pickering series using lower values of $\beta$. The velocity law obtained is
shown in Fig.~\ref{velden} (panel $a$).

The value of $\vinf$ derived from the optical lines in this work is $\sim 3$ times lower than what was derived from the observations of UV resonance lines ($\vinf \sim 1200 \kms$, \citealt{ws83}). One possible explanation for the different values of $\vinf$ derived from the UV and optical spectrum is the presence of a latitude-dependent wind, with a higher wind terminal velocity in the polar direction, and a slower equatorial wind. This possibility is further explored in Sect. \ref{2d}. 

From the optical lines, we obtained that the efficiency of the momentum transfer from the radiation to the gas ($\eta$) is much smaller in HD~45166
than in WR stars. The momentum of the gas is \mdot\vinf, while the momentum of the radiation is \lstar/c. The ratio
between the two momenta for WN stars is $\eta\simeq 2.8$ \citep{crowther07}, while for HD~45166 we obtain $\eta=0.74$ .
As a consequence, the wind driven by radiation pressure is less efficient in HD~45166 than in WR stars, which can explain the low
ratio $\vinf/\vesc$ found for HD~45166. For instance, O-type stars in about the same temperature range have $\vinf/\vesc = 2.6$ \citep{lamers95},
while WR stars have $ 1.5 < \vinf/\vesc < 4$ \citep{lamers99}. However, in HD~45166, \vesc=1320~\kms, \vinf=425~\kms, and 
therefore $\vinf/\vesc = 0.32$,  i.e, at least 5 times smaller than in typical WRs.  This is difficult to achieve for a normal wind, since it requires a fine balance between gravity and line forces. The presence of the close B7~V companion, fast rotation, and/or deviations from a spherical wind could explain such an extremely low ratio. Using \vinf=1200~\kms derived from the UV lines \citep{ws83}, the ratio 
$\vinf/\vesc$ is increased to $\sim 1$, becoming closer to the value for O-type and WR stars.

\begin{table}
\caption{Fundamental parameters derived for the qWR star in HD~45166.} 
\label{fundpar} 
\centering
\begin{tabular}{c c}
\hline\hline 
Parameter & Value\\
\hline
log (L$_\star$/L$_\odot$) & $3.75 \pm 0.08$ \\ 
T$_\star$ (K) & 70000 $\pm$ 2000\\
$\teff$ (K) & 50000 $\pm$ 2000\\
R$_\star$/R$_\odot$ ($\tau=100)$ & 0.51 \\
\reff/R$_\odot$ ($\tau=0.67)$ & 1.00 \\
$\mdot$ (\msunyr) & $2.2 \times 10^{-7}$  \\
v$_{\infty}$ (\kms) & 425 \\
$\beta$ & 4.0 \\
$f$ & 0.5 \\
v$_c$ (\kms) & 100 \\
\hline
\end{tabular}
\end{table} 

\subsection{Surface abundances \label{abund}}

The surface abundances obtained with CMFGEN are summarized in Table~\ref{abund_table}. The helium abundance was determined
adjusting the helium content in the models to match the observed intensity of \ion{He}{ii} $\lambda$ 4686, \ion{He}{ii}
$\lambda$ 5411. and the blend of \ion{He}{ii}+H$\alpha$ $\lambda$ 6560.  The value obtained is N(H)/N(He)=2.0, which
corresponds to about 2.3 times the solar abundance of He (hereafter we used the solar abundance values from
\citealt{cox2000} and references therein).

The model calculations constrained C and N abundances that are significantly higher than the abundances of O and Si. The
abundances of HD~45166 are in fact quite anomalous when compared to solar values, or to the values found in central stars of
planetary nebulae or Wolf-Rayet stars (see Table \ref{abund_comp}). This emphasizes the peculiar nature of HD~45166.

\begin{table}
\caption{Surface chemical abundances of HD~45166.} 
\label{abund_table}
\centering
\begin{tabular}{c c c c}
\hline\hline
Species & Number fraction  & Mass fraction & Z/Z$_{\odot}$\\
\hline
H 	& 	    2.0 	  & 	      3.3$\times 10^{-1}$	   & 	   0.46\\
He	 &	    1.0 	   &	      6.5$\times 10^{-1}$	    &	  2.34\\
C	 &	   3.0$\times 10^{-3}$	   &	       5.9$\times 10^{-3}$      &	  1.93   \\ 
N 	& 	   2.0$\times 10^{-3}$	  & 	      4.6$\times 10^{-3}$	   & 	 4.17\\
O 	& 	  1.5$\times 10^{-3}$	  & 	      3.9$\times 10^{-3}$	   & 	  0.41\\
Si	 &	  6.8$\times 10^{-5}$	   &	      3.6$\times 10^{-4}$	    &	  1.00\\
Fe	 &	  1.3$\times 10^{-4}$	   &	      1.4$\times 10^{-3}$	    &	  1.00\\
Ni	&	6.6$\times 10^{-6}$		  & 	    7.3$\times 10^{-5}$	   & 	1.00\\
\hline
\end{tabular}
\end{table} 

The chemical abundances must be analyzed in view of the evolutionary status of both stars in the HD~45166 system, which
is beyond the scope of this paper. Such a detailed analysis will be the subject of a forthcoming Paper III.  We anticipate
that the star is likely an exposed He core \citep{negueruela07}, which is probably related to the presence of the close
secondary. We suggest that He burning is currently going on in the nucleus, producing carbon, and, thus, explaining the He and C
overabundance. As the star still has an H-rich envelope, there certainly is a shell burning H through the CNO-cycle,
enhancing the N content. As can be seen in Table \ref{abund_table}, heavier elements such as Si, Ni and Fe have solar
abundances.

\begin{table}
\caption{Abundance ratios (in number) of HD~45166 compared to different object classes.}
\label{abund_comp}
\centerline{
\begin{tabular}{c c c c c}
\hline\hline
Object & He/H  & C/N & O/N & Reference\\
\hline
HD~45166	 & 	   0.5 	          &    1.5		 & 	 0.75   & 1 \\
Sun		  &	   0.095	  &   8.6		  &	 7.2    &  2 \\	 
PN I		 & 	   0.14 	  &   0.7--3		 & 	 1.2    &  3\\
PN II            &         0.11       &   1.4--9  &  3.5    & 3\\
WNL		 & 	   H-free	  &    0.04		 & 	 \ldots     &  4\\
WNE		  &	   H-free	  &   0.02		  &	 \ldots    &  4\\
WNEw	          &	   1.5  	  &   0.005		  &	 0.05   &  4\\
\hline
\end{tabular}}
\par \smallskip References. (1) this work; (2) \citet{grevesse07}; (3) \citet{peimbert90}; (4) \citet{crowther07}.
\end{table} 

\subsection{The ionization structure of the wind \label{ionst}}

\begin{figure} \resizebox{\hsize}{!}{\includegraphics{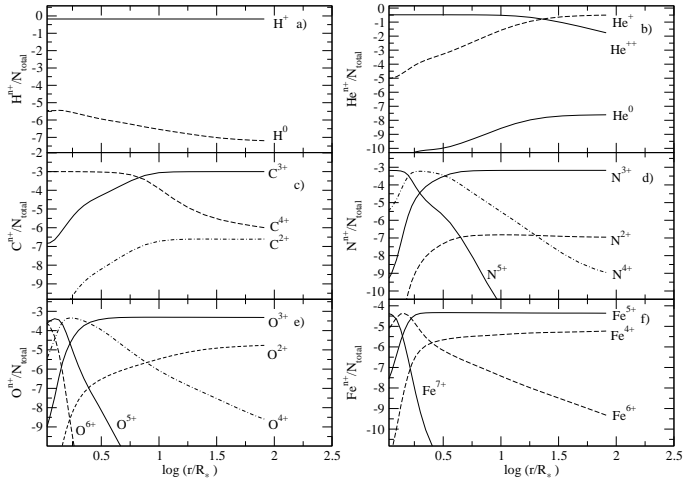}}  \caption{\label{ionstruct} Ionization fractions of HD~45166 as a function of distance, derived from the spherical CMFGEN models of the most abundant ions in the wind of the qWR star. The ionization fractions
are normalized by the total number of ions.  }
\end{figure}

Figure \ref{ionstruct} presents the ionization structure of the wind of the qWR star determined using the spherical CMFGEN model. H is
fully ionized along the whole wind, but the He, C, and Fe ionization structures are more stratified, with He$^{2+}$ and C$^{4+}$  being
more abundant in the photosphere and in the inner wind, while He$^{+}$ and C$^{3+}$ dominating at distances larger than about 10 \rstar.
This also explains the high sensitiveness of the He, C, and Fe lines on the model parameters. On the other hand, the O and N ionization 
structures are dominated respectively by O$^{3+}$ and N$^{3+}$ ions in most part of the wind -- O$^{4+}$ and N$^{4+}$ are present in
significant fraction only at distances smaller than  $\sim 1.5~\rstar $.

\begin{figure}
\resizebox{\hsize}{!}{\includegraphics{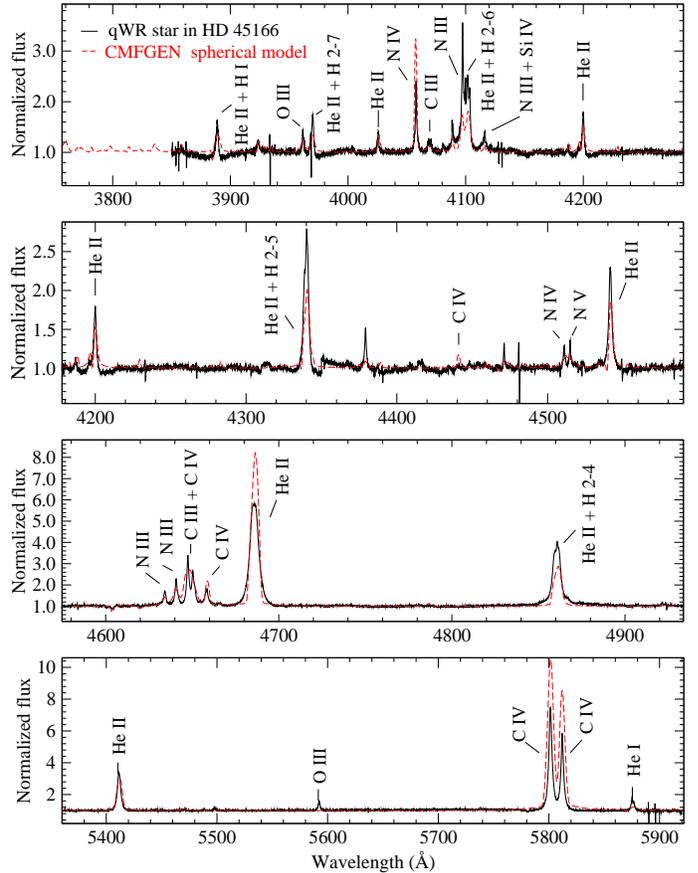}} 
\caption{\label{mod1} Comparison between the line profiles predicted by the best CMFGEN spherical model with the
observations of HD~45166, in the spectral region 3850--5950 {\AA}. The strongest spectral lines used in the analysis are identified.}
\end{figure}

\begin{figure}
\resizebox{\hsize}{!}{\includegraphics{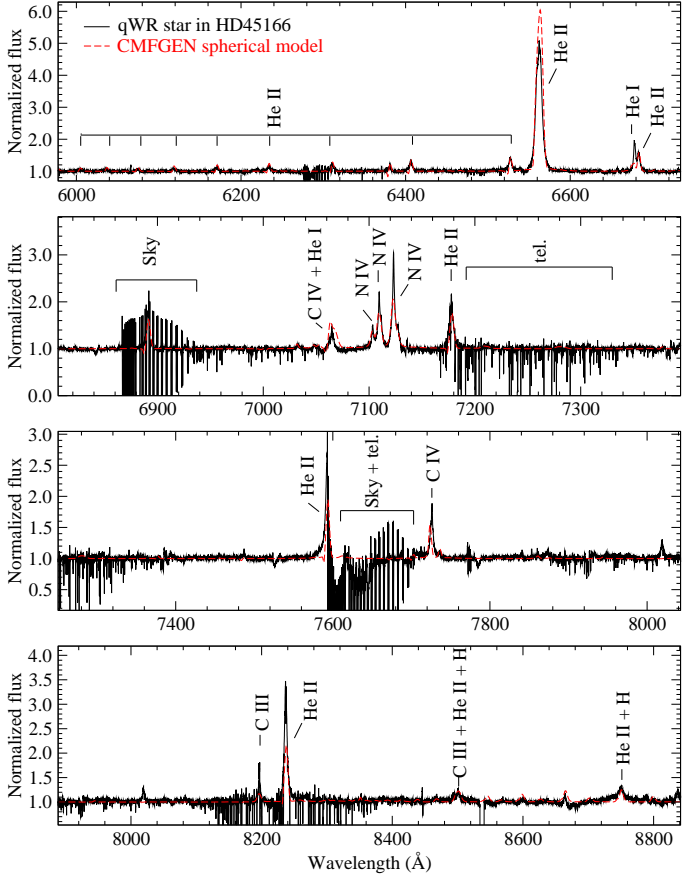}} 
\caption{\label{mod2} Same as Fig. \ref{mod1}, but for the spectral region of 6000--8800 {\AA}.}
\end{figure}

\section{\label{2d}Presence of a latitude-dependent wind}

\subsection{Observational evidence from the line profiles}

The spherical models obtained with CMFGEN reproduce the strength of most of the
emission lines, provide a superb fit to the continuum of the qWR, and a reasonable fit to the spectral lines,
considering the complex physical nature of HD~45166 (see Figs.~\ref{mod1} and \ref{mod2}). However, some discrepancies are present when comparing the
line profiles predicted by the best spherical model with the observations (Fig.~\ref{sepfitsteff}). These discrepancies might provide key 
insights on the validity of the model assumptions.

As already mentioned in Paper I, the hydrogen and helium line profiles are clearly different from the CNO lines
profiles. The later can be very well fitted by Lorentzian profiles, while \ion{He}{ii} $\lambda$ 4686 and the other
\ion{He}{ii} lines are better fitted by a Voigt/Gaussian profile. The full widths at half maximum (FWHM) are also
significantly different between the two groups of lines. 

\begin{figure}
\resizebox{\hsize}{!}{\includegraphics{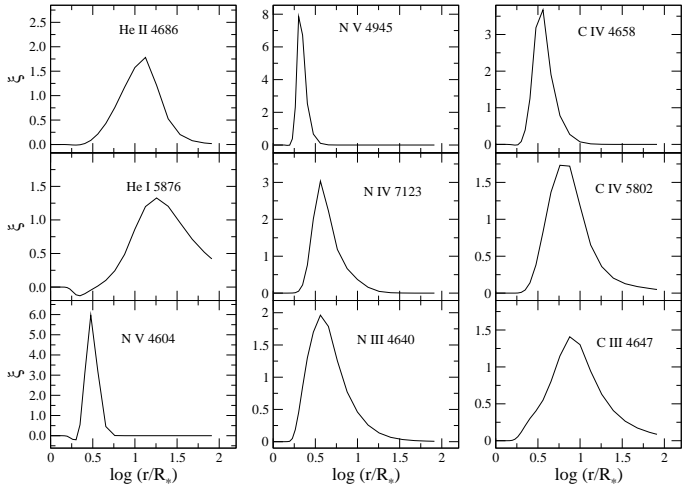}} 
\caption{\label{lineform}Line formation region for the strongest lines of HD~45166 used in the spectroscopic analysis. For
each panel the quantity $\xi$ is displayed, which is related to the EW of the line (following \citealt{hillier89}) as
$EW=\int_{\rstar}^{\infty}\xi(r)d(log\,r)$ .}
\end{figure}

Insights on the physical interpretation of the line profiles from different species can be obtained with the best CMFGEN model obtained in Sect. 
\ref{res}, which assumes a spherical wind with a monotonic
velocity law. We present in Fig.~\ref{lineform} the line formation regions for the most important diagnostic lines
present in the spectrum of HD~45166. The inner layers correspond to the formation region of higher-ionization lines such
as \ion{N}{v} and \ion{C}{iv}, while the outer layers correspond to the formation region of lower-ionization lines such
as \ion{N}{iii}, \ion{C}{iii}, and \ion{He}{i}. As can be seen in the CMFGEN model spectrum displayed in Figs. \ref{sepfitsteff}, \ref{mod1}, and \ref{mod2}, the
higher-ionization lines are predicted to be narrower than the lines of lower ionization. In addition, recombination
lines such as those from H and He are emitted in an extended region of the wind and, therefore, should be broader than
\ion{N}{iv} and \ion{N}{v} lines, for instance.

Indeed, the \ion{N}{iv} and \ion{N}{v} lines are narrow in the observations, while the \ion{He}{ii} lines are broad,
and both are well reproduced by the spherical CMFGEN models. However, the lower-ionization lines, such as \ion{C}{iii},
\ion{N}{iii}, and \ion{He}{i}, are surprisingly narrow in the observations -- much narrower than the \ion{He}{ii}
and \ion{C}{iv} lines.

We examined the effects of changing one or more physical parameters to improve the fit of the lower ionization lines, running a large number of additional CMFGEN models. None of them improved the fit to those line profiles, i.e., they did not provide narrower
lower-ionization lines. Actually, the only way to change the width of those lines using a spherical model
is using a much lower terminal velocity of $\sim 70$~\kms, which obviously did not fit the other strong spectral lines.
Moreover, it would be hard to explain the origin of such a very low wind terminal velocity in a very hot star
(\teff=50000 K). {\em In summary, a spherical symmetric wind cannot reproduce the profile of the lower-ionization lines.}

Therefore, we propose the presence of a latitude-dependent wind in the qWR star to explain the narrow profiles
observed in the \ion{N}{iii}, \ion{C}{iii}, and \ion{He}{i} lines. Using the code outlined in 
Sect. \ref{2dcalc}, we examine the effects due to changes in the wind density and terminal velocity as a function of
the stellar latitude, and present the results in Sect. \ref{2dres}.

\subsection{The 2D calculation of the emerging spectrum\label{2dcalc}}

Ideally, solving a full set of 2D radiative transfer and statistical equilibrium equations would be required to
analyze a non-spherical stellar wind as in HD~45166. However, this is currently impossible to be done taking into account
all the relevant 
physical processes, full line blanketing, and the degree of detail achieved by spherically-symmetric codes such as CMFGEN. The
large parameter space to be explored in 2D models, and the huge amount of computational effort demanded, do not allow to perform a self-consistent 2D analysis of very complex objects such as HD~45166.

Nevertheless, it is important to note that significant progresses have been achieved in developing full 2D
radiative transfer codes \citep{zsargo06,georgiev06}, which can be applied to HD~45166 in future
works. Therefore, it is desirable to explore the parameter space suitable for HD~45166, using justified assumptions to
make the computation of the observed spectrum in 2D geometry a tractable problem. 

In this work we used a recently-developed modification in CMFGEN \citep{bh05} to compute the spectrum
in 2D geometry. We refer the reader to that paper for further details about the code, whose main characteristics are
outlined below.

With the \citet{bh05} code, it is possible to examine the effects of a density enhancement and changes in the velocity
field of the wind as a function of latitude. The code uses as input the ionization structure, energy-level populations,  temperature
structure, and radiation field in the co-moving frame, as calculated by the original, spherically-symmetric CMFGEN model.
Using CMF\_FLUX \citep{hm98,bh05}, the emissivities, opacities, and specific intensity $J$
are calculated from the spherically-symmetric model.

A density enhancement and wind terminal velocity variation can then be implemented using an arbitrary
latitude-dependent density/wind terminal velocity distribution. We examined the effects due to oblate and prolate
density parameterizations, which have respectively the form
\begin{equation}
\rho_{2D} \propto \rho_{1D} (1 \mp a . cos^b \theta)\,\,,
\end{equation}
where $\theta$ is the latitude angle (0$^{\circ}$=pole, 90$^{\circ}$=equator). The changes
in the wind terminal velocity as a function of latitude were parameterized using three parameters (VB1, VB2, and VB3), 
\begin{equation}
v(r)_{2D}/v(r)_{1D}=VB1+(VB2\,\, |cos(\theta)^{VB3}|)\,\,.
\end{equation}
We assumed in this work a scaling law according to which the 2D model has the same mass-loss rate as the spherically-symmetric 
1D model.

After computing the density changes and velocity field variations, the 2D source function, emissivity, and opacity
are calculated, assuming that these quantities depend only on the new values of the scaled density. Consistent scaling laws are
used for different processes (e.g. density-squared scaling for free-free and bound-free transitions, and
linear-density scaling for electron scattering).

In the final step, the code computes the spectrum in the observer's frame, which can then be compared with the
observations.

\subsection{Model fits, results, and constraints on the wind asymmetry\label{2dres}}

\begin{figure*}
\resizebox{\hsize}{!}{\includegraphics[angle=90]{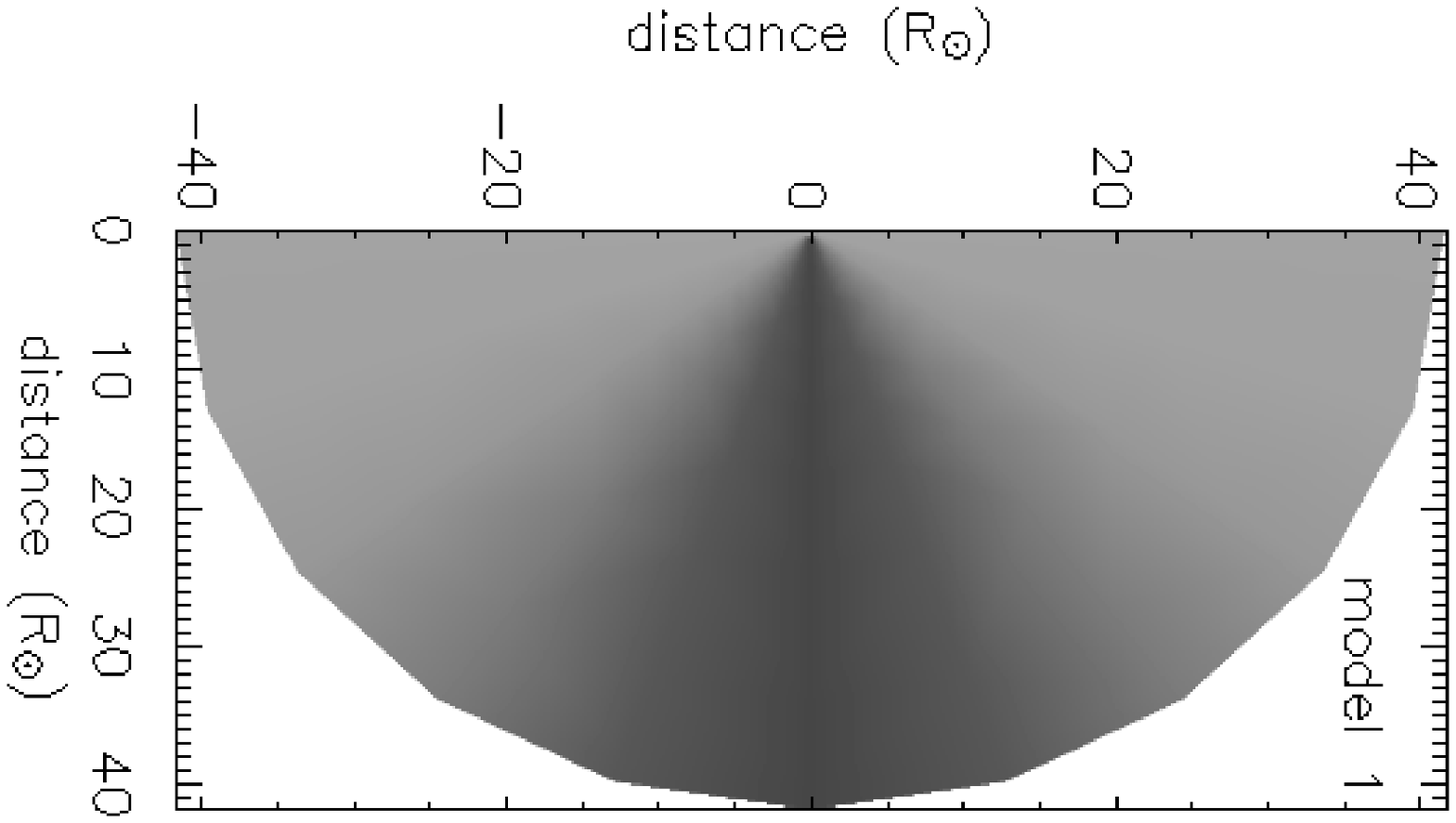}
\includegraphics[angle=90]{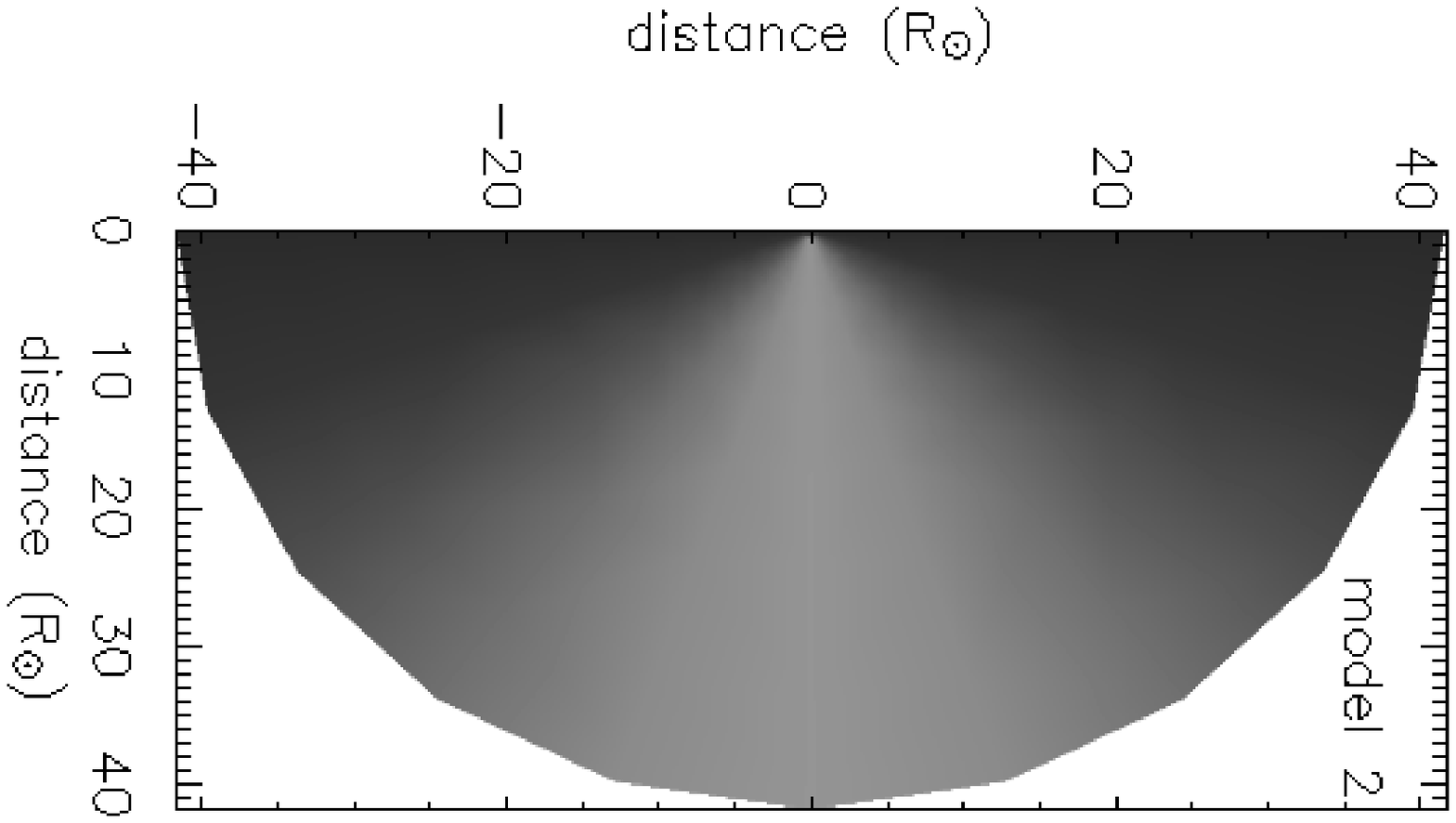}
\includegraphics[angle=90]{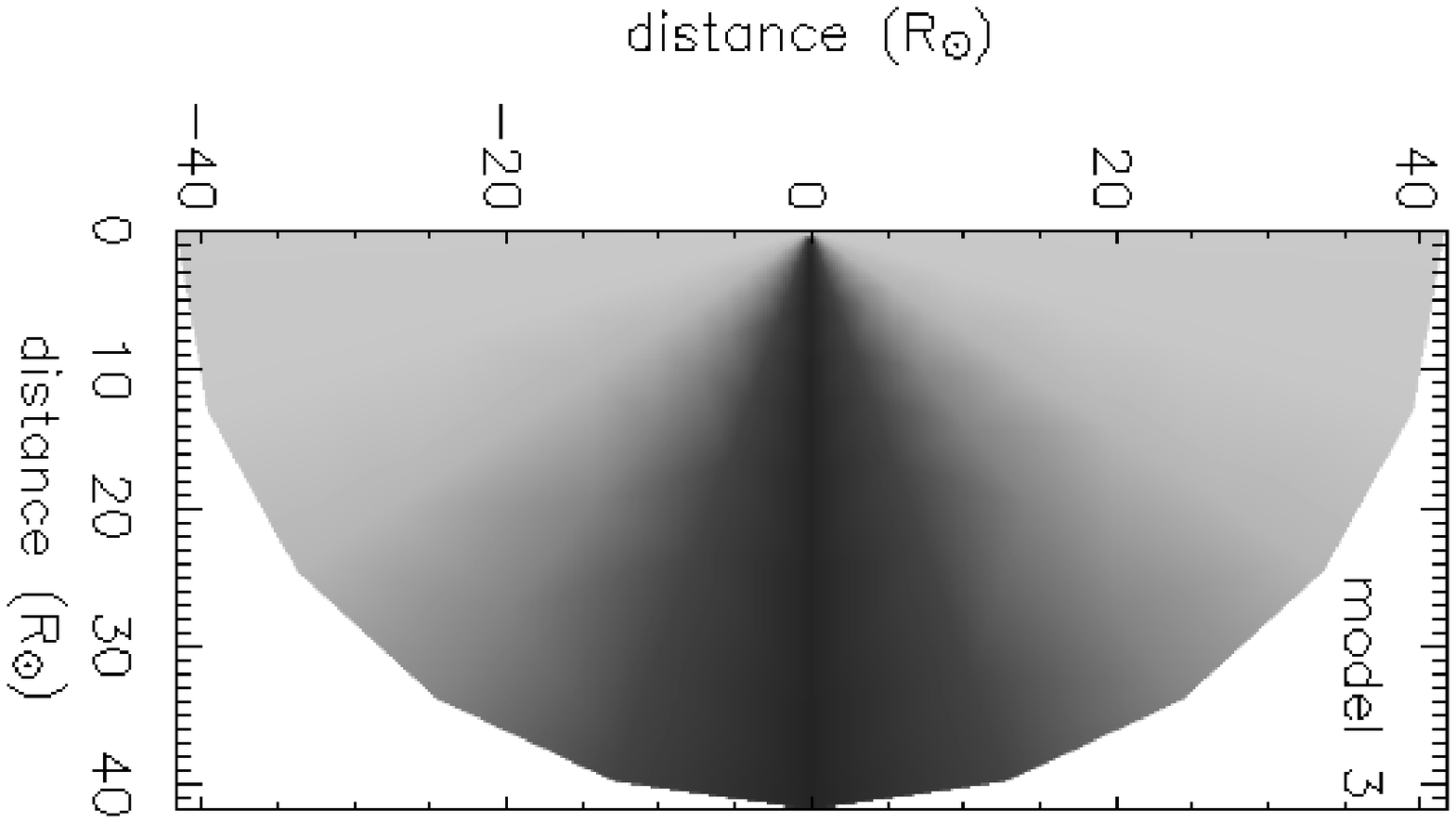}
\includegraphics[angle=90]{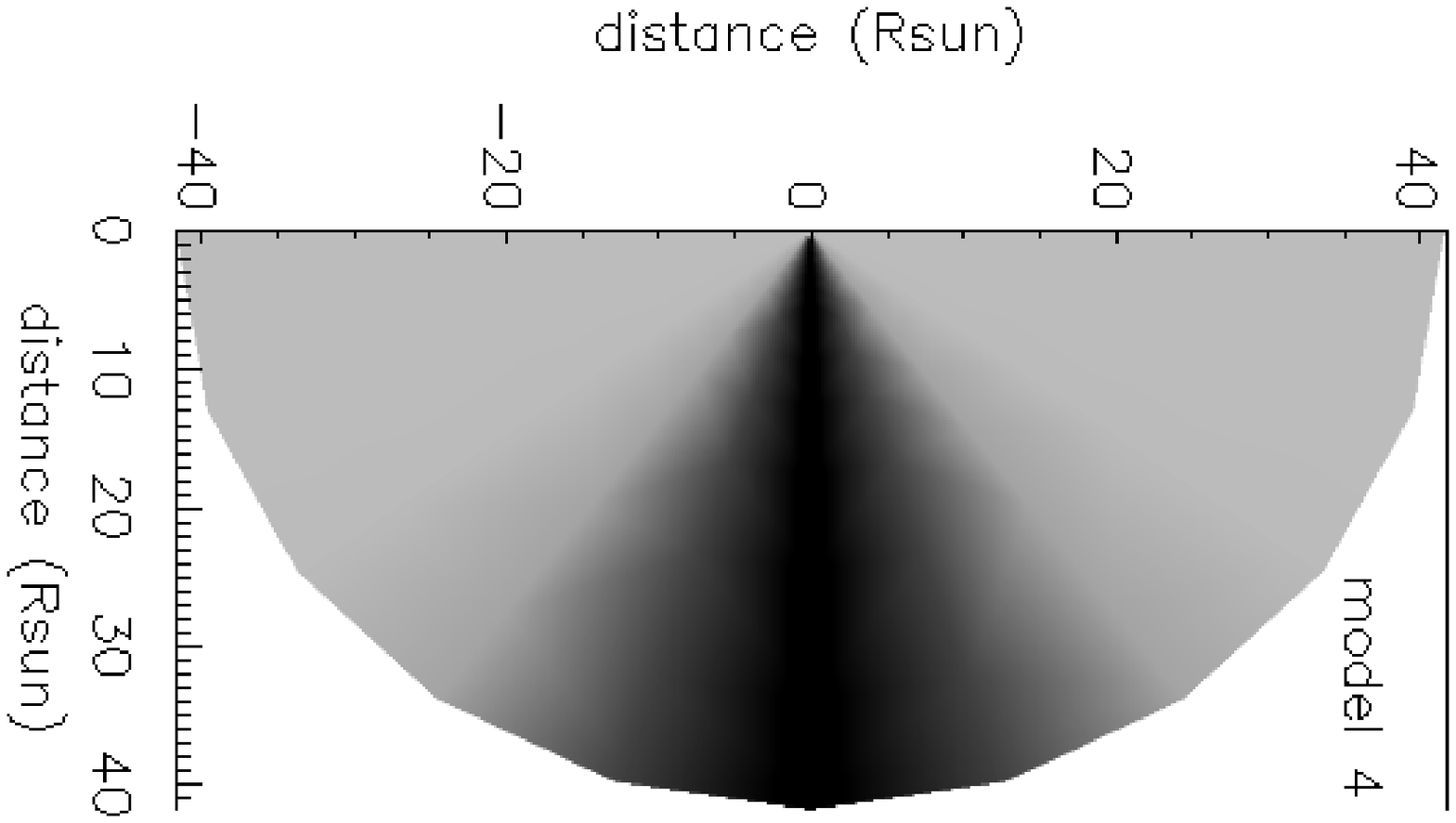}
\includegraphics[angle=90]{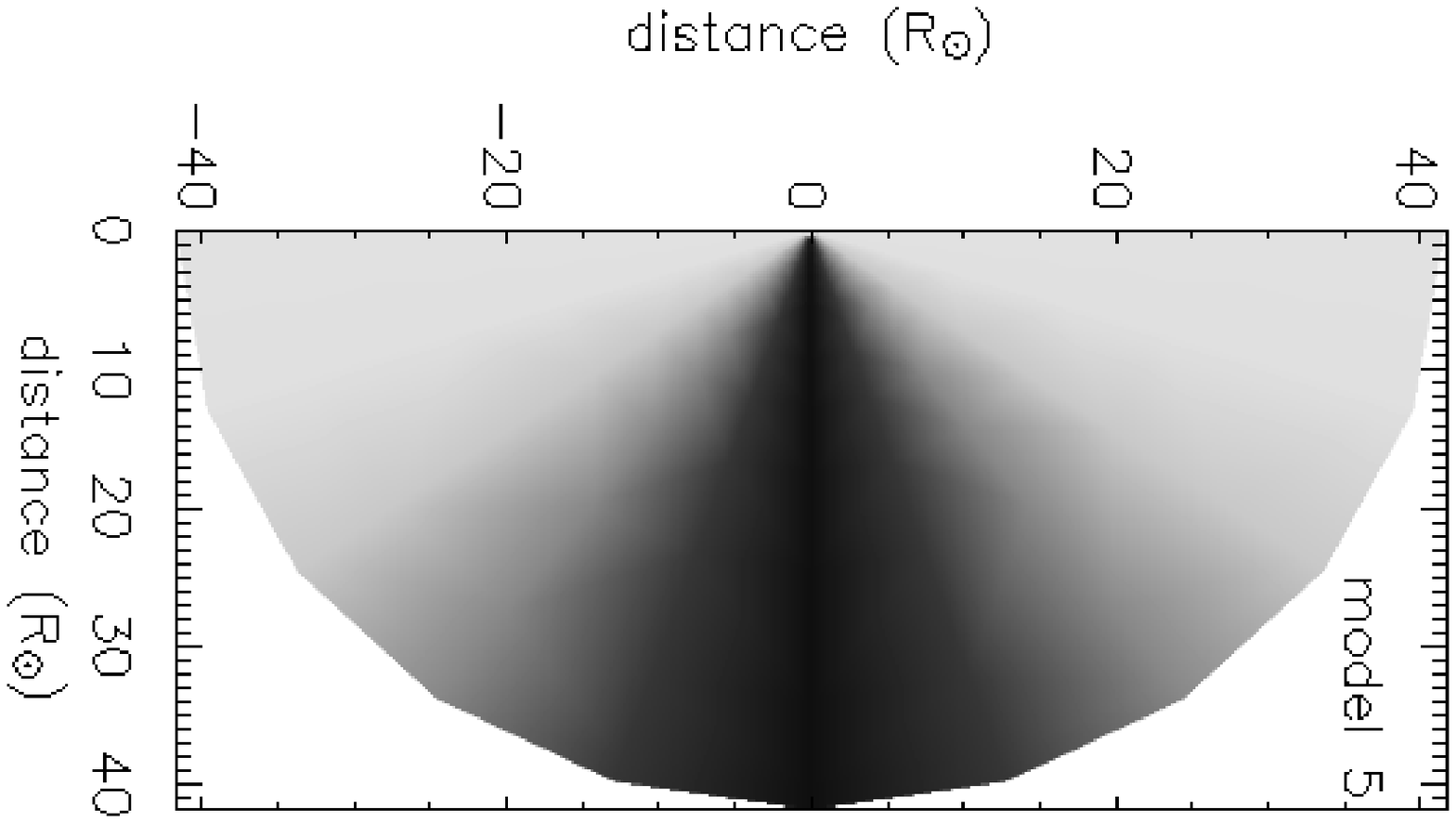}
\includegraphics[angle=90]{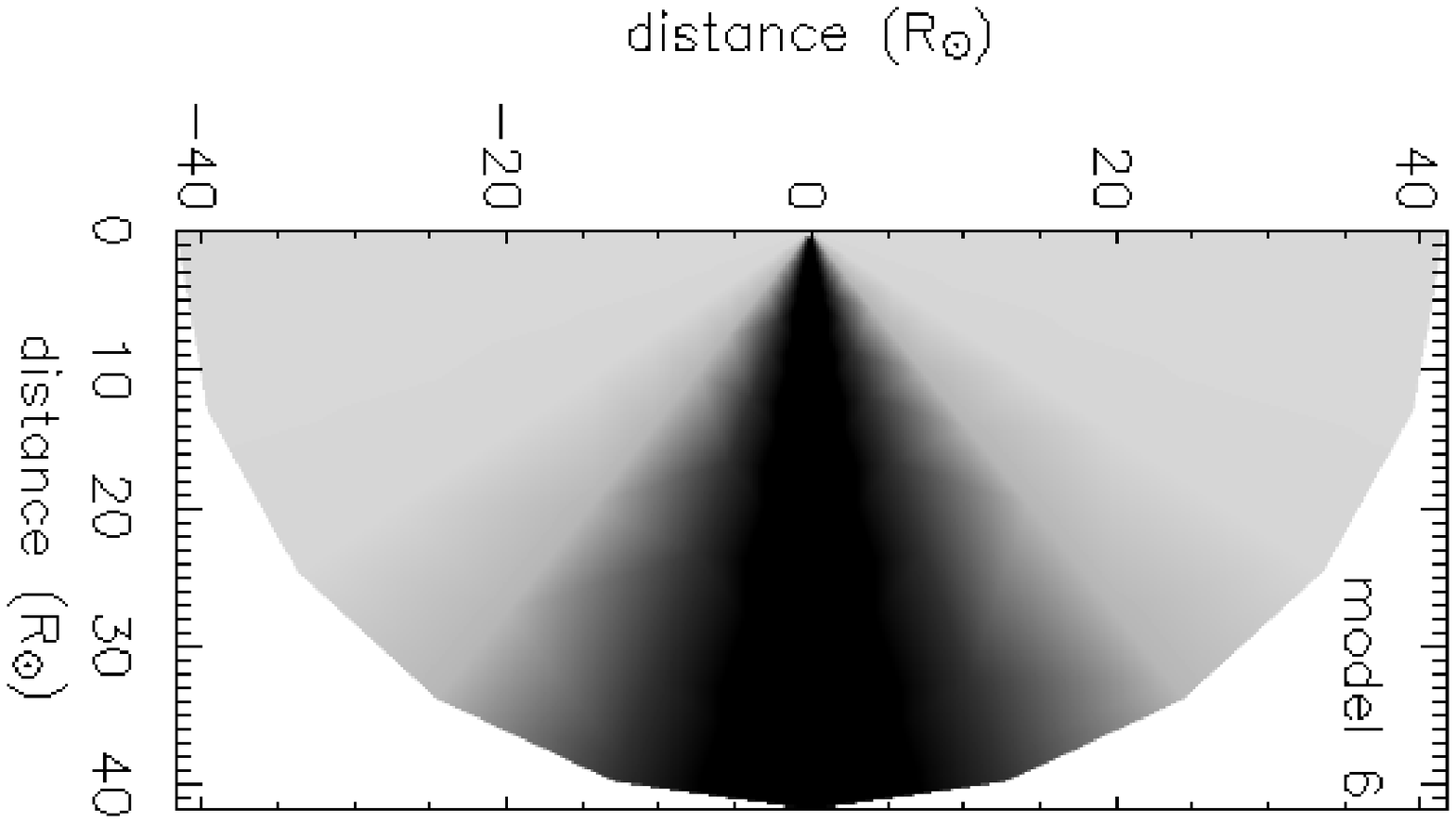}
\includegraphics[angle=90]{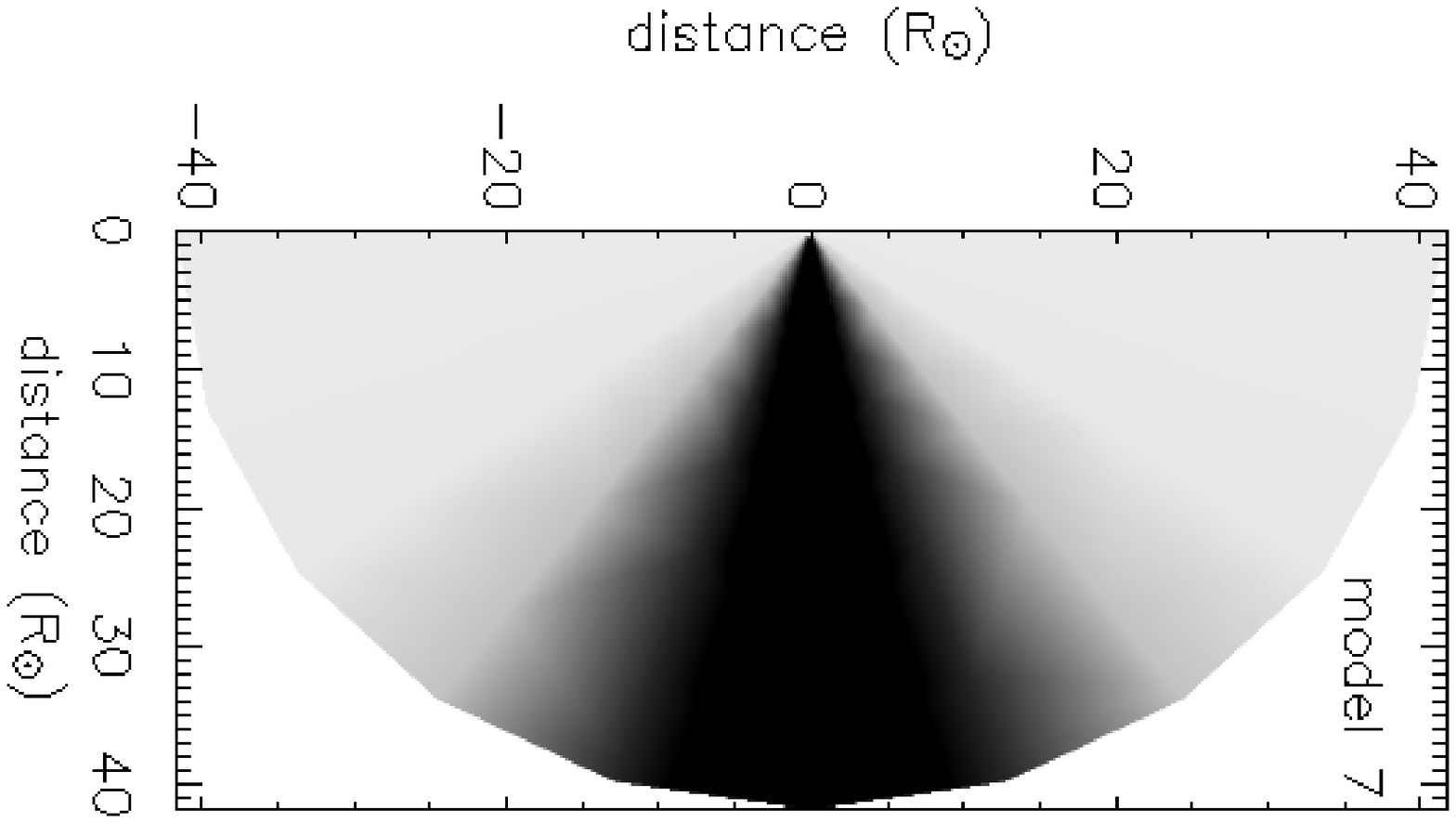}
\includegraphics[angle=90]{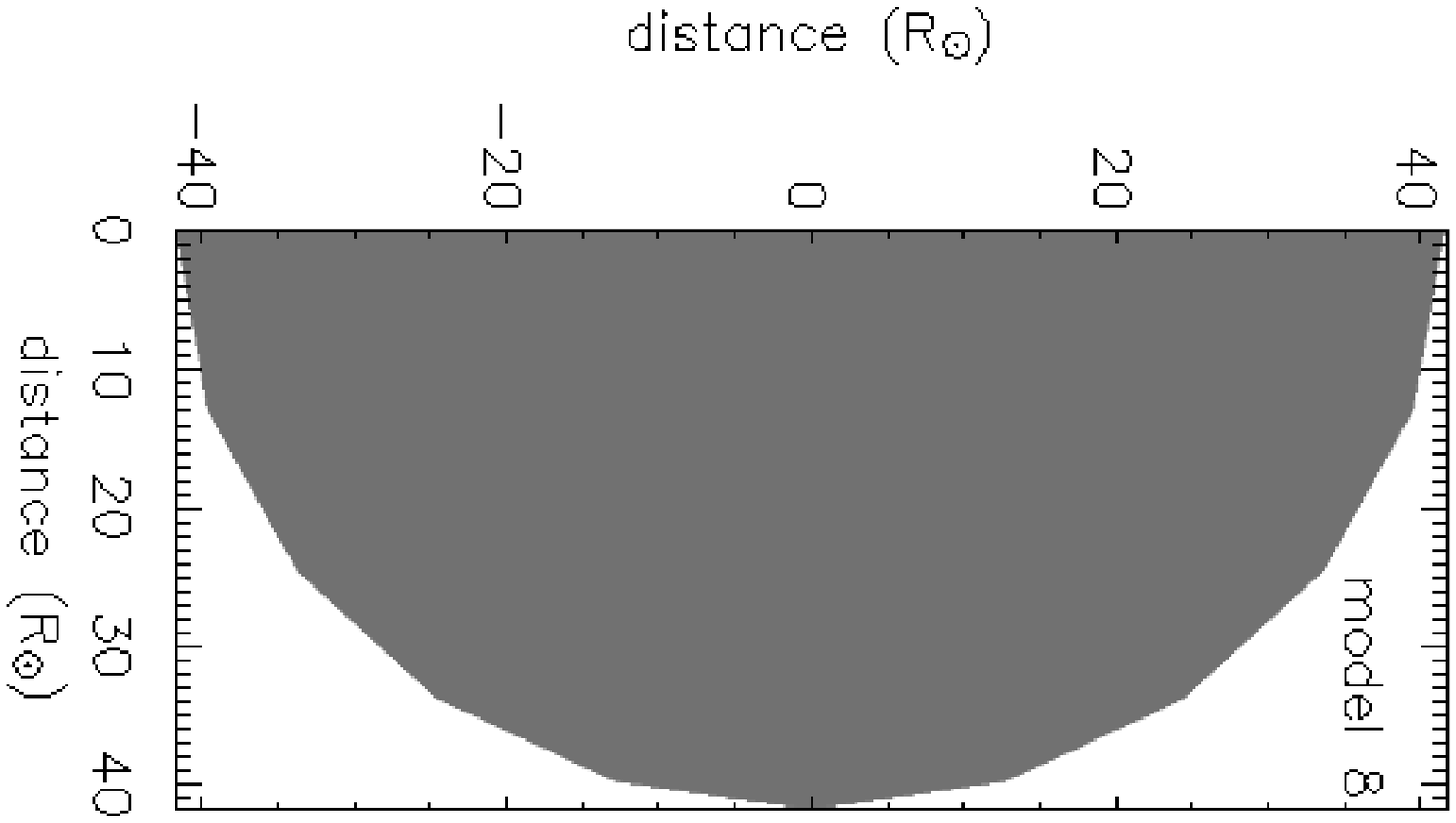}
\includegraphics[angle=90]{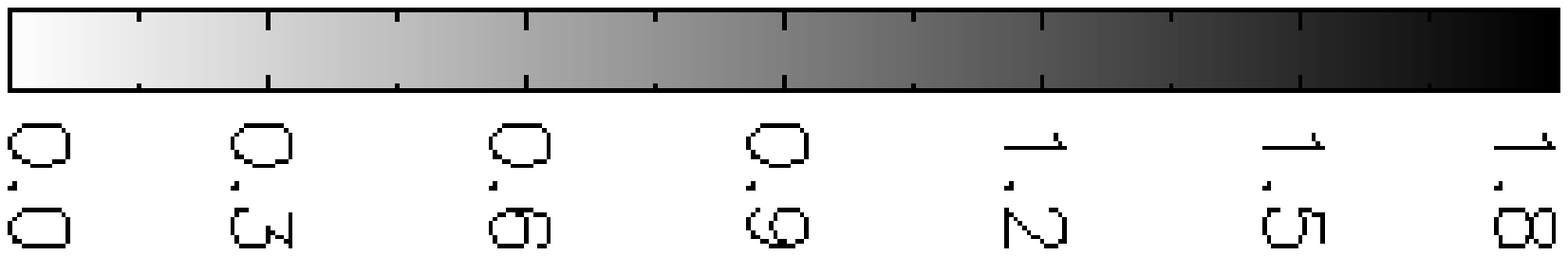}
}
\caption{\label{dencon2d} Density contrast of the HD~45166 models with latitude-dependent density distributions (normalized to
the density distribution of the spherical model), as a function of latitude. Each model is labeled as in Table \ref{2dmodel}; model 9 has the same density distribution as model 4, and is not shown. The equator of the star is at $[x,0]$, 
while the pole is located at $[0,y]$. The colorbar in the right has a linear color scale, which we adopted the same for all models,
to highlight the different density variations as a function of latitude. The maximum value in the colorbar is 1.8, which corresponds to
the maximum value found in model 5. }
\end{figure*}

The 2D model spectra were computed following the physical parameters of the best CMFGEN spherical model shown in Tables
\ref{fundpar} and \ref{abund_table}. We varied the density enhancement and velocity field as a function of latitude,
and Table \ref{2dmodel} summarizes the properties of the 2D models. Figure \ref{dencon2d} shows the latitudinal changes in the
density enhancement (normalized to the best spherical model) for the different models. We analyzed the effects of latitudinal changes 
in the density by comparing the 2D model spectra with the observed line profile of \ion{He}{i} $\lambda$ 5876, which is
very sensitive to latitudinal changes, and has the most deviating line profile in the spherical model. This line is
also isolated, minimizing the errors due to blending. Similar effects are seen in other low ionization lines of 
\ion{He}{i}, \ion{C}{iii}, and \ion{N}{iii}.

\begin{table} 
\caption{Summary of the parameters used to compute the spectrum in 2D geometry
using the \citet{bh05} code. Note that the density contrast between equator and pole is given by $1+a$ for oblate models, and $1/(1+a)$ for prolate models.} 
\label{2dmodel}
\centering
\begin{tabular}{c c c c c c c c}
\hline\hline
Model & Density  & $a$ & $b$ & Velocity & $VB1$ & $VB2$ & $VB3$ \\
& profile & & & changes? & & & \\
\hline								    
 1   & oblate	      & 1  & 2 & no	      &  \ldots  & \ldots  & \ldots   \\ 
 2   & prolate	      & 1  & 2 & no	      &  \ldots  & \ldots  & \ldots   \\ 
 3   & oblate	      & 3  & 2 & no	      &  \ldots  & \ldots  & \ldots   \\ 
 4   & oblate	      & 3  & 5 & no	      &  \ldots  & \ldots  & \ldots   \\ 
 5   & oblate	      & 7  & 2 & no	      &  \ldots  & \ldots  & \ldots   \\ 
 6   & oblate	      & 7  & 5 & no	      &  \ldots  & \ldots  & \ldots   \\ 
 7   & oblate	      & 15 & 2 & no	      &  \ldots  & \ldots  & \ldots   \\ 
 8   & none	      & \ldots  & \ldots & yes        &  0.2 & 0.8 & 4    \\ 
 9   & oblate	      & 3  & 2 & yes	      &  0.8 & 2.2 & 4    \\ 
\hline
\end{tabular}
\end{table} 

\begin{figure}
\resizebox{\hsize}{!}{\includegraphics{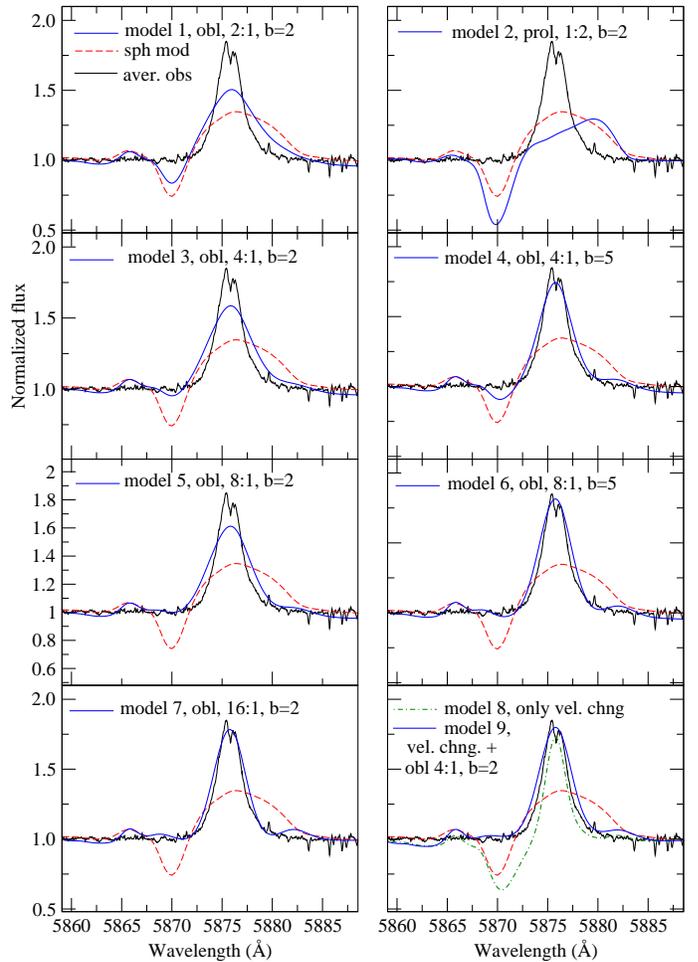}} 
\caption{\label{spec2d}Comparison between the line profiles computed with the 2D code of \citet{bh05} (blue line) with the
observations of HD~45166 around \ion{He}{i} $\lambda$ 5876 (black line). For each panel the models are labelled as in Table \ref{2dmodel}, 
followed by the density contrast equator:pole, and the parameter $b$.
The best spherical-symmetric CMFGEN model (red dashed line) is repeated in each panel for comparison (see text for further details). The weak dip in the observed \ion{He}{i} $\lambda$ 5876 line profile is likely due to incomplete subtraction of the secondary star spectrum. }
\end{figure}

Figure~\ref{spec2d} displays the observed spectrum of the qWR star around 5800~{\AA} compared with different 2D model spectra,
labeled as in Table \ref{2dmodel}. All models were computed for a viewing angle of $i=0.77 ^{\circ}$ (Paper I), and the best
spherical CMFGEN model is overplotted on each panel for comparison.

The immediate conclusion from Fig. \ref{spec2d} is that the wind of the qWR star cannot be prolate
(model 2), as the fit to the \ion{He}{i} line becomes even worse than the fit from the spherical model. This happens because, as the system
is viewed pole-on, a prolate wind implies an enhanced density in the polar regions compared to the spherical model,
producing stronger P-Cygni absorption and stronger emission at high velocities -- exactly what is seen in the spectrum of 
model 2.

We can also rule out that the narrow \ion{He}{i} line profile is only due to changes in the wind terminal velocity as a
function of latitude, as is illustrated by model 8. This model assumes a slow equatorial wind (\vinf=70 \kms) and a fast
polar wind (\vinf=700~\kms). Although the fit to the emission component is improved, the P-Cygni
absorption is still very strong, since the density enhancement has not been changed as a function of latitude.

Therefore, it seems clear that a density enhancement in the equatorial region is
required to reproduce the observed emission, and a density depletion along the polar regions is required to
reproduce the lack of a P-Cygni absorption in \ion{He}{i} $\lambda$ 5876. The minimum density contrast equator:pole that fits the
observations is 8:1, but we cannot rule out larger enhancements, such as 16:1. However, we have to keep in mind
that larger enhancements will likely change the ionization structure of the wind as a function of latitude, which
is not accounted in the current calculations. The parameter $b$, which describes how fast the density changes from
equator to pole, is quite difficult to be constrained. A model with a density contrast of 8:1 and $b=5$ (model 6)
produces almost the same He I line profile as a model with a density contrast 16:1 and $b=2$ (model 7).

We also examined changes due to both an oblate density enhancement and changes in the wind terminal velocity as a
function of latitude. However, there is no clear evidence from the fits to the optical lines that a higher velocity
polar wind is required, although it cannot be ruled out. If there is such change in the wind terminal velocity, the
minimum density contrast from equator to pole is reduced to 4:1 (see model 9 in Fig. \ref{spec2d}). Clearly, the largest effect
of a fast polar wind would be seen in the ultraviolet resonance lines.  However, we refrained from analyzing the archival IUE data simultaneously to our optical dataset due to the high variability of HD~45166 and the long time lag of 15--20 years between the IUE ultraviolet observations and our optical data.
We acknowledge the wealth of information and the fundamental importance of the UV observations to constrain the nature of HD~45166, and a forthcoming paper will be devoted to a detailed analysis of the UV spectrum.
 
It is worthwhile to stress that there is clear observational evidence of a high
velocity outflow in HD~45166 through the analysis of the discrete absorption components (DAC) in the resonance lines of HD~45166 \citep{wsh89}. However, it is also clear that marked variability is present, and at some epochs the UV resonance lines (e.g. \ion{C}{iv} $\lambda$$\lambda$ 1548--1550) do not show any high velocity absorption component \citep{wsh89}. Two questions need to be answered. 
\begin{enumerate}
\item Does the velocity of the DACs reflect the true wind terminal velocity along the polar direction? While in OB stars the maximum velocity of the DACs tend to mimic the wind terminal velocity (e.g. \citealt{ph86}), further work is required to establish whether the same holds for HD~45166. 
\item Do the DACs reflect an intrinsic property of the qWR and its steady wind or is the presence of a close companion somehow related to the variability seen in the UV lines?
\end{enumerate}
Contemporaneous ultraviolet and optical spectroscopy are required to answer these questions, and we will come back to them in Paper III.

We have not considered in this work latitudinal changes in the wind ionization structure derived from the
spherical CMFGEN modeling (Sect. \ref{ionst}), which likely occur in the case of large density changes. The inclusion of this effect
would likely reduce the density contrast determined from our analysis, but the main conclusions reached in this work would remain valid.
This is because {\it a density contrast is needed to change the ionization structure of the wind}, since the radiation field of the B7~V
companion is too weak. The exact treatment of the effects of the latitudinal changes in the ionization structure of the wind is beyond
the scope of this paper.

\section{\label{tvs}The Temporal Variance Spectrum}

We performed the Temporal Variance Spectrum (TVS) analysis in order to study the characteristics of the emission line
profiles. In this procedure, the temporal variance is calculated, for each wavelength pixel, from the residuals of the
continuum normalized spectra and the average spectrum.  For further details and discussion on the TVS method, see
\citet{fgb96}.

\begin{figure}
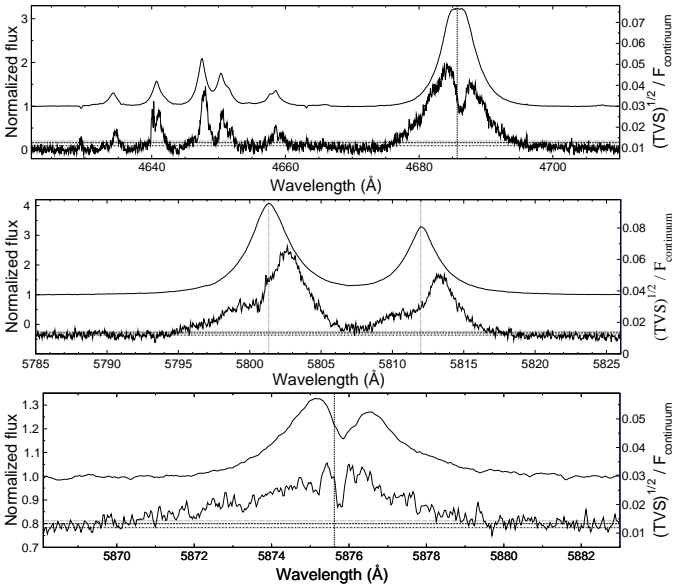

{\resizebox{0.99\columnwidth}{!}{\includegraphics[angle=0]{tvsHeII_redux.eps}} 
\resizebox{0.99\columnwidth}{!}{\includegraphics[angle=0]{tvsCIV_redux.eps}}
\resizebox{0.99\columnwidth}{!}{\includegraphics[angle=0]{tvsHeI_redux.eps}}}
\caption{\label{tvsfigs} Average intensity spectrum and TVS of HD~45166 for \ion{He}{ii} $\lambda$ 4686 (upper panel), \ion{C}{iv} $\lambda\lambda$
5801--5812 (middle), and  \ion{He}{i} $\lambda$ 5876 (bottom). The TVS statistical threshold significance of $p$=1\%, 5\%, and 30\% are represented
by the dotted, dashed, and dot-dashed horizontal lines, respectively. The vertical dotted lines mark the position of the
zero velocity for the relevant line presented in each panel. See text for details.}
\end{figure}

We calculate the TVS for the strongest spectral lines, such as \ion{He}{ii} $\lambda$ 4686, \ion{He}{i} $\lambda$ 5876, and
\ion{C}{iv} $\lambda$$\lambda$ 5801--5811 (Fig. \ref{tvsfigs}). The TVS contain a large amount of information
that, in spite of not being always of immediate interpretation, certainly can be helpful in understanding the
wind structure. Initially, one should notice that, as well as in the line profiles, also in the TVS the
spectrum is quite different when we compare, for instance, \ion{He}{ii} $\lambda$ 4686 and \ion{C}{iv}
$\lambda$$\lambda$ 5801--5812. At the wings of \ion{He}{ii} $\lambda$ 4686, the TVS is close to a Lorentzian profile. 
At smaller velocities, however, the TVS profile of the line deviates more and more from a Lorentzian profile until it
reaches a local minimum. Assuming that the Lorentzian profile is a signature of optically-thin line
emission, the wind seems to become transparent for \ion{He}{ii} line photons at velocities larger than 270~\kms. 
The corresponding radius is $r = 10 \rstar$, according to the velocity law determined using the best CMFGEN
model. The CMFGEN model itself predicts that $\tau(\ion{He}{ii} 4686)<1$ for $r>15$~\rstar (see Fig. \ref{velden}).

On the other hand, the TVS of the \ion{He}{i} $\lambda$ 5876 line presents a peculiar behavior. While the 
intensity profile is asymmetric to the red, the TVS profile is asymmetric to the blue. This discrepancy is
quite strong and its interpretation is not trivial. The stronger H lines of H$\alpha$ and H$\beta$ are
blended to the \ion{He}{ii} lines and, therefore, their profile and interpretation is more complex. 

A noticeable feature in the TVS  profiles is the presence of a central dip in the \ion{He}{ii} and \ion{He}{i}
lines. In the case of \ion{He}{ii}, this dip is centered at a velocity of $+35.3 \pm 0.6$~\kms. We do not
have an interpretation for this velocity. In the TVS profile of \ion{He}{i} $\lambda$ 5876, however, the dip
is at a different velocity of  $+4.1 \pm  0.1$~\kms, which is consistent with the systemic velocity of $+4.5
\pm  0.2$~\kms~(Paper I). On the other hand, in the intensity spectrum of \ion{He}{i} $\lambda$ 5876, the absorption
is centered at $+9.2 \pm 0.3$~\kms. Interestingly, this is remarkably similar to the
photospheric velocity found for the absorption lines of the secondary star. Therefore, we suggest that such dip in the TVS of \ion{He}{i} $\lambda$ 5876 is due to the orbital motion of the secondary star.

Most of the intensity line profiles present in the spectrum of HD~45166 are quite symmetric.
The lines of \ion{C}{iv} $\lambda\lambda$ 5801--5811, for instance, are very well fitted by symmetrical Lorentzian profiles, as already mentioned in Paper I. However, in the TVS the asymmetries are much more evident.
For the lines of \ion{N}{iii} $\lambda$$\lambda$ 4634--4640, \ion{C}{iii} $\lambda$$\lambda$ 4647--4650 and 
\ion{C}{iv} $\lambda$ 4658, as well as for \ion{C}{iv} $\lambda$$\lambda$ 5801--5812 (see Fig. \ref{tvsfigs}), the TVS
profiles are very different from the intensity profiles of the respective lines as well as from the TVS of the
\ion{He}{ii} $\lambda$ 4686 line. While in the TVS of the latter the blue wing is 30\% more intense than the red wing, in
the lines of C and N the opposite happens. For instance, in the TVS of \ion{C}{iv} $\lambda$$\lambda$ 5801--5812, the red
wings are 3 times stronger than the blue wings (see Fig. \ref{tvsfigs}). We interpret this phenomenon as evidence for 
temporal changes in the gas density of the equatorial wind region. This produces a variable optical depth (due to absorption or electron
scattering) and primarily affects the red part of the profile, which is formed in the receding side of the wind. 
The nature of the variability in the gas density is not clear, and a detailed interpretation is beyond the scope of this paper.
Two possible interpretations are variability in the (equatorial) mass-loss rate, or debris from a hypothetical, variable mass transfer
from the secondary star. The latter hypothesis will be examined in Paper III.

A characteristic indicator of each line is the ratio between its variance and the intensity, $\sigma/I$.
Table \ref{sigmai} presents this ratio for the strongest lines of HD~45166. \footnote{The telluric lines in
absorption have a much higher TVS in our data (28\%), and, therefore, those lines can be easily identified and
differentiated from the stellar lines. On the other hand, the interstellar lines should not appear in the
spectrum of the TVS. The TVS can be a useful method to distinguish between the several types of lines
present in a rich emission and absorption line spectrum such as in HD~45166.} It is interesting to notice that
\ion{He}{i} lines have the highest $\sigma/I$ values (2.6\%), followed by \ion{H}{i}+\ion{He}{ii} lines, which are
estimated to be about 2.5\%. The higher variability of low ionization lines
can be interpreted in the context of a Wind Compression Zone (WCZ) scenario \citep{icb96}, in
which a lower ionization region is produced at the equator of the system (see Sect. \ref{wcz}). 
Variability in the WCZ may cause a higher $\sigma/I$ ratio for the \ion{He}{i} and \ion{H}{i}
lines when compared to the higher ionization lines of \ion{He}{ii} or \ion{C}{iv}. It is not clear whether the same scenario can explain the variability detected in the photospheric UV lines of \ion{Fe}{v}, \ion{He}{II}, and \ion{N}{iv} \citep{wsh89}.

\begin{table}
\caption{$\sigma/I$ for the strongest lines present in HD~45166. Both $\sigma$ and $I$ were measured at the centroid of the lines.}  
\label{sigmai} 
\centering
\begin{tabular}{c c c}
\hline\hline 
Ion & Line & $\sigma/I$ (\%)\\
\hline
\ion{H}{i}+\ion{He}{ii}   &  4861  &    2.0\\
\ion{H}{i}+\ion{He}{ii}   &  6563  &    2.2\\
\ion{He}{i}		  &  5876  &    2.6\\
\ion{He}{i}		  &  6678  &    2.1\\
\ion{He}{ii}		  &  4541  &    1.6\\
\ion{He}{ii} 		  &  4686  &    1.5\\
\ion{He}{ii}		  &  6527  &    1.4\\
\ion{He}{ii}		  &  6683  &    1.3\\
\ion{C}{iv}		  &  5801  &    1.6\\
\ion{C}{iv}		  &  5812  &    1.5\\
\ion{C}{iii}		  &  4647  &    1.9\\
\ion{N}{iii}		  &  4510  &    1.1\\
\ion{N}{iii}		  &  4640  &    1.9\\
\hline
\end{tabular}
\end{table} 

\section{\label{discussion}Discussion}

\subsection{Gravitational redshift and the line formation regions}

The gravitational redshift is measurable in the CNO emission lines of HD~45166 (see Paper I) . The 
CNO\,{\sc III} lines have an average radial velocity of $5.0 \pm 0.8$~\kms~(Table \ref{fwhmvelr}). This is compatible with the
systemic velocity $\gamma=4.5 \pm 0.2$~\kms~(Paper I). However, the \ion{N}{iv}--\ion{N}{v} lines 
have an average radial velocity of $8.3 \pm 1.2$~\kms~(Table \ref{fwhmvelr}). These higher-ionization lines should be emitted
in the wind inner layers and, therefore, should be redshifted when compared to the lines of lower ionization. Indeed, the 
observed velocity shift of those more ionized species, compared to the less ionized ones, is $3.3 \pm 1.2$~\kms. 
This means that, taking the systemic velocity as a reference, the CNO\,{\sc III} lines should be emitted at distances
larger than $\sim 10$~\rstar, assuming the velocity law derived in Section
\ref{vinf}, and $\mathrm{M}_{qWR}=4.2$~\msun~(Paper I). Interestingly, those distances are comparable to the size of the Roche lobe of the qWR star.
In contrast, the gravitational redshift of the \ion{N}{iv}--\ion{N}{v} lines indicates that they should be emitted close to the hot star, at distances of the order of
$1.5-2.0$~\rstar. 

\begin{figure}
\resizebox{\hsize}{!}{\includegraphics{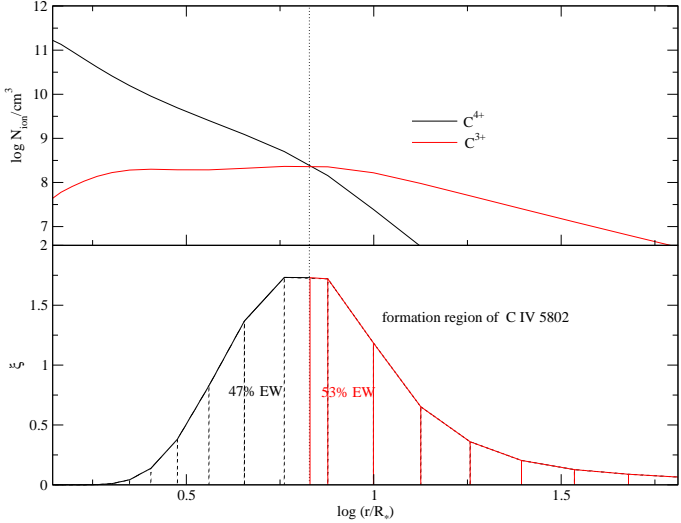}} 
\caption{\label{civfluor}Upper panel: radial density profile of the C$^{3+}$ and C$^{4+}$ ions along the wind of HD~45166. Lower panel:
line formation region of \ion{C}{iv} $\lambda\lambda$ 5801. The quantity $\xi$ is related to the EW of the line (following
\citealt{hillier89}) as
$EW=\int_{\rstar}^{\infty}\xi(r)d(log\,r)$ .}
\end{figure}

The above values are in reasonable agreement with the line formation regions presented in Fig. \ref{lineform}, taking 
into account that the lines actually form in an extended wind region. One should also consider that the emission lines
can be a result of pure recombination (e.\,g. \ion{C}{iv} $\lambda$ 4658) or, at least partially, due to continuum fluorescence
(e.g. \ion{C}{iv} $\lambda$$\lambda$ 5801--5812). This is shown in Fig. \ref{civfluor}, where the line formation region of \ion{C}{iv}
$\lambda$$\lambda$ 5801 is compared to the radial variation of the density of C$^{3+}$ and C$^{4+}$ ions. It can be seen that roughly 50\% 
of the line emission come from an inner region where the dominant ionization stage of carbon is C$^{4+}$, suggesting that the line is formed
by recombination. However, a significant fraction of the line formation occurs under very low C$^{4+}$ density conditions, making it very unlikely that
recombination is the main mechanism of line emission in those regions. Instead, we suggest that the line is emitted due to continuum fluorescence of the 
C$^{3+}$ ions.

This discussion illustrates that the physical conditions found in HD~45166 is far too complex to be properly described by a spherical wind model. Therefore, understanding the line formation processes and determining  the line formation regions provide invaluable information on the 
nature of HD~45166.

\begin{table}
\caption{Average FHWM and radial velocity of the spectral lines for the most abundant ions present in HD~45166.}  
\label{fwhmvelr} 
\centering
\begin{tabular}{c c c}
\hline\hline 
Ion & $<FWHM>$ & $<v_r>$\\
& (\kms) & (\kms)\\
\hline
\ion{C}{iv}		  &  142 $\pm$ 19 & 4.4 $\pm$ 0.4   \\
\ion{C}{iii}		  &  91 $\pm$ 6   & 5.4 $\pm$ 1.5  \\
\ion{N}{v}		  &  79 $\pm$ 20  & 7.3 $\pm$ 1.6   \\
\ion{N}{iv}		  &  82 $\pm$ 11  & 8.6 $\pm$ 1.5  \\
\ion{N}{iii}		  &  96 $\pm$ 4   & 4.0 $\pm$ 1.1   \\
\ion{O}{iii}		  &  113 $\pm$ 9  & 5.0 $\pm$ 1.6  \\
$<\ion{CNO}{iii}>$	  &  102 $\pm$ 4  & 5.0 $\pm$ 0.8  \\
$<\ion{N}{v}+\ion{N}{iv}>$  &  78 $\pm$ 10  & 8.3 $\pm$ 1.2  \\
\hline
\end{tabular}
\end{table} 

\subsection{Rotation, wind asymmetry, and formation of a Wind Compression Zone \label{wcz}}

Depending on the rotation velocity of the star and on the velocity law of the wind, a wind compression disc (WCD, \citealt{bc93}), or a wind
compression zone (WCZ, \citealt{icb96}), can be formed. The most fundamental parameter to consider is the ratio of the rotational velocity of the
star (\vrot) to the critical break-up velocity (\vcrit). Since HD~45166 is seen almost pole-on ($i=0.77 ^{\circ}$, Paper I), it is
impossible to detect the rotational broadening of lines formed close to the photosphere, and to further determine the rotational velocity using the
\citet{bh05} code, similar to what has been done, for instance, for the luminous blue variable AG Carinae \citep{ghd06}.

Nevertheless, the effects due to the rotation of HD~45166 can be quantified in terms of the ratio of the critical period for break-up (\pcrit) to the
rotational period (\prot),
\begin{equation}
\vrot/\vcrit=\pcrit/\prot\,\,,
\end{equation} where $\vcrit = (G\mstar/\rstar)^{1/2}$. For values of $\vrot/\vcrit$ close
to 1 a WCD can be formed \citep{bc93}, while for moderate rotational velocities a WCZ can be produced \citep{icb96}. 

Using the stellar parameters determined for HD~45166 (see Sect. \ref{res}), we have $\pcrit \simeq 1.0$ h. Therefore, unless the 
rotational period is close to 1 hour, the rotation is too slow to cause the formation of a WCD. However, the formation of a WCZ may still be possible, 
and might explain the degree of asymmetry detected in the wind of HD~45166 (see Sect. \ref{2d}). In Paper III we will show
evidence of a period of 2.4 hours, which might be related to rotational modulation.

The difference between the two scenarios outlined above is that, in the case of a WCD, the formation of a stationary shock
heats up the disc, and compresses it to a thin sheet with high density \citep{bc93}. In the case
of a WCZ, this shock does not exist and the compression area is much thicker and less dense \citep{icb96}.
While for a WCD the ratio between equatorial and polar density is of the order of $\sim 1000$, for a WCZ with
the parameters of HD~45166 this ratio is $\sim5-10$. The value of the density contrast derived from our 2D modeling 
is 4--8 (see Sect. \ref{2d}), which is of the same order as predicted by a WCZ scenario. 

In addition, in the case of a WCD, the shock formed in the equatorial region should be hot enough to be detected in soft X-rays. However, HD~45166 
was not detected by the {\it ROSAT} X-ray satellite. Superionization due to X-rays \citep{bc93} should also produce enhanced \ion{N}{v} and
\ion{O}{vi} emission near the equator, which is not observed. In contrast, on the equator there is enhanced emission of low ionization species
such as \ion{He}{i} lines (see Sect. \ref{2d}), which argues against the presence of a WCD.

However, there are at least two caveats which might change the WCZ/WCD scenario and have to be taken into account in future works of HD~45166. 
First, non-radial line forces might play an important role in shaping the 2D wind structure \citep{go00}. Secondly, if the star is rapidly rotating, the effective temperature is also going to vary as a function of latitude \citep{vonzeipel24}, and the 2D wind structure will likely be affected. 

\section{\label{conc}Conclusions}

In the following, we present the main conclusions of this paper.

\begin{enumerate}

\item We used the radiative transfer code CMFGEN to quantitatively analyze the wind of the qWR star HD~45166, and
to constrain its fundamental parameters. Comparing the model spectrum with  high-resolution optical
observations, it was possible to determine the effective temperature, radius, luminosity, mass-loss rate, wind
terminal velocity, and surface abundances. The temperature of the hot star is $\teff=50000 \pm 2000$~K, 
the radius is $\reff = 1.00$~\rsun, and the luminosity is $\mathrm{log} (L/\lsun) = 3.75 \pm 0.08$.

\item The wind parameters are quite anomalous. The mass loss rate determined is $\mdot
= 2.2 \times 10^{-7}$~\msunyr, which is about 4 times higher than the value proposed by \citet{vb78}. 
Analyzing the optical lines, we determined the terminal velocity of the wind to be \vinf=425~\kms, and not 1200~\kms as measured by \citet{wsh89}, since
the optical spectrum shows no evidence of such a high-velocity wind.
While for O-type and WR stars $\vinf/\vesc>1.5$, in the case of HD~45166 it is much only 0.32. In addition, 
the efficiency of momentum transfer from the radiation field to the wind is $\eta=0.74$, which is at least a factor of
4 lower than what is found for WR stars.

\item The star is helium-rich, with N$_{\ion{H}{}}$/N$_{\ion{He}{}}$ = 2.0, obtained by fitting simultaneously the \ion{He}{ii} lines which are
unblended, such as \ion{He}{ii} $\lambda$ 4686, and those blended with hydrogen Balmer lines, such as
 \ion{He}{ii}+H$\alpha$ $\lambda$ 6560. The He content is consistent with the analysis of the He II Pickering
lines, which shows that a significant abundance of hydrogen exists in the wind. The helium content of HD~45166
is actually 4 times higher than the abundance determined by \citet{vb78}. The CNO abundances are quite
peculiar, and very different from those of the Sun, of central stars of planetary nebulae, as well as
from WN stars.

\item The wind of the qWR star is not spherically symmetric. Analyzing the observed line profile by comparison with
model spectra computed in 2D geometry, we were able to infer the presence of an oblate wind, with a minimum
density contrast of 8:1 from equator to pole. Taking into account our assumptions, 
higher density contrasts, for instance 16:1, cannot be discarded since they yield similar line profiles.

\item We also analyzed whether the wind terminal velocity changes as a function of latitude, but there is no clear
evidence that this is the case. If we include a fast polar wind (\vinf=1300 \kms) and a slower equatorial wind
(\vinf=300 \kms), the observed line profiles can be fitted with a smaller density contrast from equator to pole of 4:1. Therefore, such possibility might be a promising way to reconcile the values of $\vinf$ obtained from optical and UV analyses.

\item We applied the Temporal Variance Spectrum (TVS) technique to the strongest emission lines of HD~45166. The TVS
profiles of the lines is quite complex. The ratio sigma/intensity varies from 1.5\% for \ion{He}{ii} lines up to 2.5\%
for those of \ion{He}{i}. Both \ion{He}{i} and \ion{He}{ii} lines also present a central dip in their TVS, 
in spite of this signature not being present in the intensity spectrum of \ion{He}{ii}. 

\item The TVS of some emission lines, especially \ion{C}{iv}, show that variable absorption or scattering occurs near the
equator; as a consequence, the red wings are more variable than the blue wings of the same line. The cause of this
variability could be blobs of gas moving out in the wind compression zone or debris from variable mass transfer
from the secondary star. 

\item Differential gravitational redshift seems to be noticeable in the velocities of high-ionization lines when
compared to the velocities of low-ionization lines. 

\item The latitudinal changes in the density and velocity might be explained by the formation of a Wind Compression Zone \citep{icb96},
which would imply that the qWR star has a relatively high rotational velocity.

\end{enumerate}

\begin{acknowledgements}

We are grateful to the referee Dr. Allan Willis for the constructive comments and suggestions to improve the paper.
It is a pleasure to thank John Hillier and Joe Busche for making the CMFGEN and the \citet{bh05} codes available, and for continuous support on the codes. We are also grateful to Thomas Driebe, John Hillier, Florentin Millour, and Janos Zsarg\'o for the careful reading and detailed comments on the original manuscript.
J. H. Groh acknowledges financial support from the Max-Planck-Gesellschaft (MPG), and Brazilian agencies FAPESP (grant 02/11446-5) and CNPq
(grant 200984/2004-7).

\end{acknowledgements}

\bibliography{refs_hd45166}

\appendix

\section{The H/He abundance and the He II Pickering decrement}

The hydrogen abundance determined by \citet{vb78} is N$_{\rm H}$/N$_{\rm He} \sim 5.3$ (by number), which is about 4 times higher than the value
derived through the detailed radiative transfer modeling using CMFGEN (N$_{\rm H}$/N$_{\rm He}$=2.0 by number, see Sect. \ref{abund}). This
difference is significant, and we will analyze the \ion{He}{ii} Pickering decrement to verify this discrepancy. 

\begin{figure}[!h]
\resizebox{0.9\columnwidth}{!}{\includegraphics{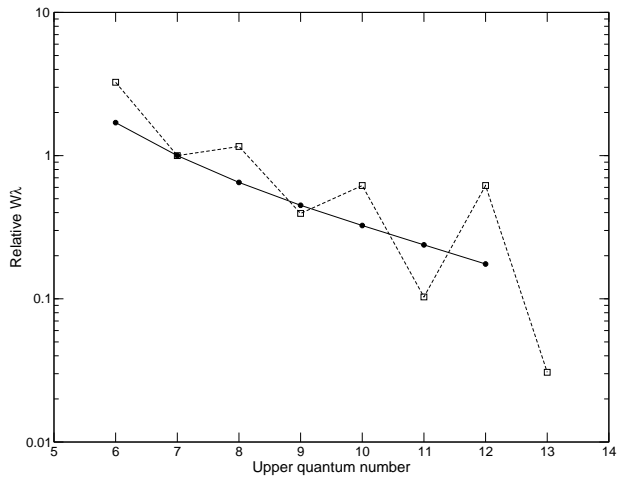}}
\vspace{0.0cm} 
\caption{\label{pick} Intensity of the \ion{He}{ii} lines of the Pickering series as a function of the upper quantum number, normalized 
such as that I(\ion{He}{ii} $\lambda$ 5411) = 1. The open squares represent the observed values of HD~45166, while the filled circles correspond
to the theoretical prediction for an optically-thin gas \citep{ost89}, using tabulated values from \citet{os04}.} 
\end{figure}

Although the hot star of HD~45166 is He-rich (~65\% in mass, at least in the wind), the presence of hydrogen is an important
aspect to be carefully considered in the analysis. As mentioned by \citet{ws83}, judging from the \ion{He}{ii} Pickering 
decrement, hydrogen is definitely present. To check this, we compared our observations with the expected theoretical values. 
For practical reasons we normalized the predicted and the observed values to 1 for \ion{He}{ii} $\lambda$ 5411. 
Figure \ref{pick} shows the comparison between the two series, and we clearly see the typical oscillation pattern that appears when hydrogen is 
present. It is also possible to infer that the intensities of the lines that coincide with hydrogen Balmer lines are about twice as
strong as the expected ones if there was no Balmer emission.  One way of analyzing the presence of hydrogen in a helium-rich wind spectrum, following \citet{os04}, is by defining the Pickering parameter, $p$:
\begin{equation}
p=\frac{I(4859+4861)}{[I(4541).I(5411)]^{0.5}}\,\,.
\end{equation}

Using the predicted theoretical intensities that one expects for a pure \ion{He}{ii} spectrum, calculated for a low-density
gas, $p = 0.97 \pm 0.01$ \citep{ost89}. Any value of $p$ larger than 1 would mean hydrogen is present in the wind. 
In the case of HD~45166, the measured value is $p \simeq 2.95$. In case both H and He lines were optically thin, it
would be easy to determine the relative abundance between the two species. In this case, given the specific emissivity
of the respective transitions \citep{ost89},
\begin{equation}
\label{thin}
\frac{N(H{+})}{N(He^{++})}=  p - 1\,\,.
\end{equation}

For HD~45166, one gets $N(H^{+})/N(He^{++}) \simeq 1.95$. The largest uncertainty in this value is the hypothesis
that all of the involved lines are optically thin. This may be a good approximation at the wings of the lines but not
at their cores. For an optically thick case we have \citep{conti83}:
\begin{equation}
\label{thick}
\frac{N(H{+})}{N(He^{++})}=  p^{1.5} - 1\,\,.\,
\end{equation}
and thus we obtain $N(H^{+})/N(He^{++}) \simeq 4.06$.

\section{The CNO emission line table for HD~45166}

\input{cno_table.tex}

\end{document}

%% file: cno_table.tex
\begin{longtable}{c c c c c c c c}
\caption{\label{cno_table} Equivalent width of the CNO emission lines in HD
45166 obtained from the observations. References: Willis \& Stickland (1983) for ultraviolet lines (1200--3300 {\AA}), and this work for optical lines (3700--9000 {\AA}).
Key: y=line is present; y?=doubtful
identification due to low S/N; ?=doubtful identification due to presence of a
telluric line; bl=blended with other lines; n=line is definitely not present.} \\
\hline\hline
& & \multicolumn{2}{c}{O} &\multicolumn{2}{c}{N} & \multicolumn{2}{c}{C}\\
\cline{3-4} \cline{5-6} \cline{7-8} 
Term & J--J &	$\lambda$ ({\AA})	& W$_{\lambda }$ ({\AA})&	$\lambda$ ({\AA})	& W$_{\lambda }$ ({\AA})&	$\lambda$ ({\AA})	& W$_{\lambda }$ ({\AA})\\
\hline
\endfirsthead
\caption{continued.}\\
\hline\hline
& & \multicolumn{2}{c}{O} &\multicolumn{2}{c}{N} & \multicolumn{2}{c}{C}\\
\cline{3-4} \cline{5-6} \cline{7-8} 
Term & J--J &	$\lambda$ ({\AA})	& W$_{\lambda }$ ({\AA})&	$\lambda$ ({\AA})	& W$_{\lambda }$ ({\AA})&	$\lambda$ ({\AA})	& W$_{\lambda }$ ({\AA})\\
\hline
\endhead
\hline
\endfoot
& & \multicolumn{2}{c}{\ion{O}{iii}} & & & & \\
\hline
\multicolumn{2}{c}{IP(eV)}&\multicolumn{2}{c}{54.94} & &  & &\\					
\multicolumn{2}{c}{{\it Singlets}}&  & &&&&\\
1P$^0$--1S              &       1--0     &	1247  & ?       & & & &\\
3s 1P$^0$--3p 1P	&	1--1	&	5592  &	0.75	& & & & \\
3p 1D--3d 1F$^0$	&	2--3	&	3962  &	0.29    & & & & \\
4p 1S--5s1P$^0$	&	0--1	&	5268  &	0.13    & & & & \\
4p 1D--5s 1P$^0$ 	&		&	4569  &	0.05    & & & & \\
3s 1P$^0$--3p 1D	&		&	2984  &	0.88?   & & & & \\

\multicolumn{2}{c}{{\it Triplets}}&  & &&&&\\
3P$^0$--3P      &      		&       1175  & 1.86 ? & & & & \\
3s 3P$^0$--3p 3D&       2--3	&	3760  &	1.4    & & & & \\
	        &       1--2	&	3755  &	0.99   & & & & \\
		&       0--1	&	3757  & 1.1    & & & & \\	
		&       2--2	&	3791  & \ldots	       & & & & \\	
	        &       1--1	&	3774  &	\ldots       & & & & \\	
3p 3P--3d 3D$^0$&	2--3	&	3715  &	0.99   & & & & \\
  		&       1--2	&	3707  & 0.35   & & & & \\	
		&       1--1	&	3703  & \ldots       & & & & \\
3s 3P--3p 3D$^0$& 	2--3	& 	4081  & 0.12   & & & & \\
		& 		& 	4074  & \ldots       & & & & \\
4d 3P$^0$--3s 3D&	1--2	&	7455  & 0.08   & & & & \\
		&		&	7475  &\ldots        & & & & \\
		&		&	7482  & \ldots       & & & & \\
		&		&	7492  & \ldots       & & & & \\
\hline
& & \multicolumn{2}{c}{\ion{O}{iv}} & \multicolumn{2}{c}{\ion{N}{iii}} & & \\
\hline
\multicolumn{2}{c}{IP(eV)}&\multicolumn{2}{c}{77.41} & \multicolumn{2}{c}{47.45}& &\\					

2p2 2P--2p3 2Do  &	 3/2--5/2    &	  1344  & \ldots	  &	    1752  &  \ldots       &   & \\
  	        &	  1/2--3/2   &    1339  &  1.82   &	    1748  & \ldots	   &   & \\
	   	&	  3/2--3/2   &	  1343  &  0.92   &	    1751  & \ldots	   &   & \\
3s 2S--3p 2P$^0$    &	  1/2--3/2   &	  3063  & \ldots	  &	    4097  &  1.66  &   & \\
 	        &	  1/2--1/2   &	  3072  & \ldots	  &	    4103  &  0.84  &   & \\
3s 2P$^0$--3p 2D    &	 3/2--5/2    &	  3349  & \ldots	  &	    4200  &  0.77  &   & \\
         	&	  1/2--3/2   &	  3348  & \ldots	  &	    4196  &  0.16  &   & \\	 
	  	&	 3/2--3/2    &	  3378  & \ldots	  &	    4216  & \ldots	   &   & \\
 3p 2P--3d 2P$^0$   &	 3/2--3/2    &	  2133  & \ldots	  &	    2984  &  0.88  &   & \\
		&	  1/2--3/2   &	  2127  & \ldots	  &	    2979  & \ldots	   &   & \\
		&	 3/2--1/2    &	  2126  & \ldots	  &	    2977  & \ldots	   &   & \\
                &	  1/2--1/2   &	  2120  & \ldots	  &	    2973  & \ldots	   &   & \\
3p 2P$^0$--3d 2D    &	 3/2--5/2    &	  3412  & \ldots	  &	    4641  &  1.37  &   & \\
 	        &	  1/2--3/2   &	  3404  & \ldots	  &	    4634  &  0.75  &   & \\	
  		&	 3/2--3/2    &	  3414  & \ldots	  &	    4642  &	?  &   & \\
4d 2D--5f 2F$^0$	&	 5/2--7/2    &	  3563  &\ldots	  &	    4004  &  0.07  &   & \\
            		&	 3/2--5/2    &	  3560  &\ldots	  &	    3999  & \ldots	   &   & \\
4f 2F$^0$--5g 2G	&		     &	 	&	  &	    4379  &  0.56  &   & \\
3d 2P$^0$--6d 2D	&		     &	 	&	  &	    8494  &  y	   &   & \\
5f 2F$^0$--6g 2G	&		     &	 	&	  &	    8019  &  0.54  &   & \\
3s 4P$^0$--3p 4D   &	 5/2--7/2    &	  3386  &\ldots	  &	    4515  &  0.19: &   & \\
	 	&	 1/2--3/2    &	  3381  &\ldots	  &	    4511  &  0.21  &   & \\
3p 4D--3d 4F$^0$   &	 7/2--9/2    &	  3737  &\ldots	  &	    4867  &  0.20  &   & \\
	     	&	 5/2--7/2    &	  3729  &\ldots	  &	    4861  &\ldots	   &   & \\
		&	 3/2--5/2    &	  3725  &\ldots	  &	    4859  &\ldots	   &   & \\
\hline	
& & \multicolumn{2}{c}{\ion{O}{v}} & \multicolumn{2}{c}{\ion{N}{iv}} & \multicolumn{2}{c}{\ion{C}{iii}} \\
\hline		
\multicolumn{2}{c}{IP(eV)}&\multicolumn{2}{c}{113.90} & \multicolumn{2}{c}{77.47}&\multicolumn{2}{c}{47.89} \\							
\multicolumn{2}{c}{{\it Singlets}}&  & &&&&\\
2p 1P$^0$--2p2 1D	 &	  1--2	 &	    1371  &   2.71 &		1719 &    4.97  &    2297  &   1.88   \\
2p 1P$^0$--2p2 1S	 &	  1--0	 &	    774   &\ldots	   &		955  &	\ldots        &    1247  &   y      \\
3s 1S--3s 1P$^0$	 &	  0--1	 &		  &	   &		1188 &    0.19  &    1591  &	\ldots      \\
3s 1S--3p 1P$^0$	 &	  0--1	 &	    5114  &   y?   &	  	6381 &    0.21: &    8500  &   y      \\
3p 1P$^0$--3d 1D	 &	  1--2	 &	    3145  &\ldots	   &		4058 &    1.11  &    5695  &   0.096  \\
3d 1D--3s 1P$^0$	 &		 &		  &	   &		2279 &  \ldots        &    2982  &   y?     \\
3d 1D--3d 1F$^0$	 &	  2--3	 &		  &	   &		1310 &    0.26  &    1541  &\ldots	      \\
3d 1D--5f 1F$^0$	 &	  2--3	 &	    509   &\ldots	   &		1297 &    y?    &    1382  &\ldots	      \\
4p 1P$^0$--5d 1D	 &	  1--	 &	    1419  &   0.30 &		     &	        &  	   &	      \\
4f 1F$^0$--5g 1G 	 &	  3--4	 &	    1708  &\ldots	   &		3078 &  \ldots        &    4187  &   0.127  \\
\multicolumn{2}{c}{{\it Intercombination}}&  & &&&&\\
2s2 1S--2p 3P$^0$   &	0--	 &	    1218 &\ldots	  &		1486 &	0.11	&	1909&	n      \\
\multicolumn{2}{c}{{\it Triplets}}&  & &&&&\\
2p 3P$^0$--2p2 3P	 &	 1--0	&	  759	&\ldots	    &	    923  &\ldots		 &    1175 &   1.86    \\
			 &	 	&	  759	&\ldots	    &	 	 &	 	 &    1172 &	\ldots       \\
3s 3S--3p 3P$^0$	 &	 1--2	&	  2781  &\ldots	    &	    3479 &\ldots	 	 &    4647 &   2.38    \\
		 	 &	 1--1	&	  2787  &\ldots	    &	    3482 &\ldots	 	 &    4650 &   1.46    \\
			 &	 1--0	&	  2790  &\ldots	    &	    3484 &\ldots	 	 &    4651 &   0.42:   \\
3s 3P$^0$--3p 3P	 &	 2--2	&	  2731  &\ldots	    &	    3463 &   y  	 &    4666 &   y       \\
			 &	 	&	  2744  &\ldots	    &	    3461 &   y  	 &    4673 &   ?       \\
			 &	 	&	  2729  &\ldots	    &	    3475 &   y  	 &    4663 &   ?       \\
3s 3P$^0$--3p 3D	 &	 2--3	&	  4124  &\ldots	    &	    5204 &   y  	 &    6744 &   0.17    \\
			 &	 	&	  4120  &\ldots	    &	    5200 &   y  	 &    6731 &   0.077:  \\
			 &	 	&	  4125  &\ldots	    &	    5205 &   y? 	 &    6727 &\ldots	       \\
3p 3P$^0$--3p 3D	 &	 	&	  1055  &\ldots	    &	    1272 &   0.92	 &    1578 &\ldots	       \\
		 	&	 	&	  1058	&\ldots	    &		 &		 &	   &	       \\
		 	&	 	&	  1059	&\ldots	    &		 &		 &	   &	       \\
		 	&	 	&	  1060	&\ldots	    &		 &		 &	   &	       \\
		 	&	 	&	  1061	&\ldots	    &		 &		 &	   &	       \\
3p 3P$^0$--3d 3D	 &	 2--3	&	  5598  &  n	    &	    7123 &   3.31	 &    9715 &\ldots	       \\
			 &	 1--2	&	  5580  &\ldots	    &	    7109 &   1.91:	 &    9705 &\ldots	       \\
		 	&	 0--1	&	  5573  &\ldots	    &	    7103 &   0.92	 &    9701 &\ldots	       \\
		 	&	 2--2	&	  5604  &\ldots	    &	    7127 &   0.70:	 &    9718 & \ldots          \\
		 	&	 1--1	&	  5583  &\ldots	    &	    7111 &   y  	 &    9706 &\ldots	       \\
		 	&	 2--1	&	  5607  &\ldots	    &	    7129 &\ldots	 	 &    9719 &\ldots	       \\
3d 3D--3d 3F$^0$	 &	 	&		&	    &	    1326 &   0.14    	 &	   &	       \\
3d 3D--5f 3F$^0$	 &	 	&		&	    &	 	 &	 	 &    1296 &\ldots	       \\
4s 3S--4p 3P$^0$	 &	 1--2	&	  7437  &  n	    &	    6220 &   y  	 &    7796 &   n       \\
			 &	 1--1	&	  7443  &  n	    &	    6215 &   y  	 &    7780 &   n       \\
		 	&	 1--0	&	  7458  &\ldots	    &	    6212 &   y? 	 &    7772 &\ldots	       \\
4p 3P$^0$--5d 3D	 &	 2--	&	  1418  &  0.61     &	    2036 &\ldots	 	 &    3610 &\ldots	       \\
4d 3D--5f 3F$^0$	 &	 	&	  1507  &  0.37     &	    2318 &\ldots	 	 &    3889 &   y?      \\
5d 3D--6f 3F$^0$ 	 &	 	&	  2757  &\ldots	    &	 	 &	 	 &    7487 &   0.134   \\
4f 3F$^0$--5g 3G	 &	 4--5	&	  1644  &\ldots	    &	    2648 &\ldots	 	 &    4068 &   y       \\
		 	 &	 	&		&	    &	    2647 &\ldots	 	 &    4070 &   y       \\
		 	 &	 	&		&	    &	    2646 &\ldots		 &	   &	       \\
5f 3F$^0$--6g 3G	 &	 	&	  2707  &\ldots	    &	    4707 &   y  	 &    7487 &   y       \\
5g 3G--6h 3H$^0$	 &	 	&	  2942  &\ldots	    &	    4606 &   0.077	 &    8196 &   y       \\
6h 3H$^0$--7i 3I	 &	       	&	  4930  &   y	    &	    7703 &   y  	 &    13696&\ldots	       \\

\hline 
& & \multicolumn{2}{c}{} & \multicolumn{2}{c}{\ion{N}{v}} & \multicolumn{2}{c}{\ion{C}{iv}} \\
\hline		
\multicolumn{2}{c}{IP(eV)}&\multicolumn{2}{c}{} & \multicolumn{2}{c}{97.89}&\multicolumn{2}{c}{64.49} \\							
\multicolumn{2}{c}{{\it Doublets}}&  & &&&&\\	
2s 2S-- 2p 2P$^0$	&	 1/2--3/2  &&&	  1239  &  9.0      &	    1548  &  1.34  \\
			&	 1/2--1/2  &&&	  1243  &\ldots	    &	    1551  &  0.55  \\	       
3s 2S--3p 2P$^0$	&	 1/2--3/2  &&&	  4604  &  abs      &	    5801  &  16.68 \\	       
			&	 1/2--1/2  &&&	  4620  &  abs      &	    5812  &  9.96  \\
4s 2S--4p 2P$^0$	&	 1/2--3/2  &&&	  11331 &\ldots	    &	    14335 &\ldots	   \\
			&	 1/2--1/2  &&&	  11374 &\ldots	    &	    14362 &\ldots	   \\
5s 2S--5p 2P$^0$	&	 1/2--3/2  &&&	  22572 &\ldots	    &	    28617 &\ldots	   \\
			&	 1/2--1/2  &&&	  22654 &\ldots	    &	    28675 &\ldots	   \\
\multicolumn{2}{c}{{\it Multiplets}}&  & &&&&\\					   
(4--5)		&		&&&	    1620   & 0.21      &      2530   &  0.26 \\
		&		&&&	    1622   & 0.15      &	     &	     \\
		&		&&&	    1655   &\ldots	       &	     &	     \\
(5--6)		&		&&&	    2981   &\ldots	       &      4658   & 0.42: \\
(5--7)		&		&&&	    1860   &\ldots	       &      2907   &\ldots	     \\
(6--7)		&		&&&	    4945   & 0.103     &      7726   & 1.55  \\  
(6--8)		&		&&&	    2998   &\ldots	       &      4685   & bl    \\
(7--8)		&		&&&	    7618   &\ldots	       &      1908   &	\ldots     \\
(7--9)		&		&&&	    4520   &\ldots	       &      7063   & y     \\
(8--9)		&		&&&	    10980  &\ldots	       &      17371  &\ldots	     \\
(7--10)		&		&&&	    3502   &\ldots	       &      5471   & y     \\
(8--10)		&		&&&	    6478   &\ldots	       &      10124  &\ldots	     \\
(9--10)		&		&&&	    15536  &\ldots	       &      24278  &\ldots	     \\
(8--11)		&		&&&	    4943:  &\ldots	       &      7737   & y     \\
(9--11)		&		&&&	    8927   &\ldots	       &      13954  &\ldots	     \\
(10--11)        &		&&&	    21000  &\ldots	       &      32808  &\ldots	     \\
(8--12)		&		&&&	    4191   &\ldots	       &      6560   & bl    \\
(9--12)		&		&&&	    6747   &\ldots	       &      10543  &\ldots	     \\
(10--12)	&		&&&	    11928  &\ldots	       &      18635  &\ldots	     \\
(11--12)	&		&&&	    27593  &\ldots	       &      43138  & \ldots      \\  
(9--13)		&		&&&	    5670   &\ldots	       &      8858   &\ldots	     \\
(10--13)	&		&&&	    8918   &\ldots	       &      13946  & \ldots      \\   	
\end{longtable}